\documentclass[manuscript,screen,nonacm]{acmart}

\makeatletter
\@twosidefalse      
\@mparswitchfalse   
\renewcommand\@titlefont{\fontsize{14}{16}\selectfont\bfseries} 
\makeatother

\usepackage[utf8]{inputenc}
\usepackage{svg}
\usepackage{subcaption}
\usepackage{lscape}
\usepackage{float}

\AtBeginDocument{%
  }

\renewcommand\footnotetextcopyrightpermission[1]{} 
\begin{document}

\title[Exploring Social Media Clones]{Social Media Clones: Exploring the Impact of Social Delegation with AI Clones through a Design Workbook Study}

\author{Jackie Liu}
\affiliation{%
  \institution{University of British Columbia}
  \city{Vancouver}
  \country{Canada}
}
\email{anjieliu@cs.ubc.ca}

\author{Mehrnoosh Sadat Shirvani}
\affiliation{%
  \institution{University of British Columbia}
  \city{Vancouver}
  \country{Canada}
}
\email{mehrshi@cs.ubc.ca}

\author{Hwajung Hong}
\affiliation{%
  \institution{Korea Advanced Institute of Science \& Technology}
  \city{Daejeon}
  \country{Korea}
}
\email{hwajung@kaist.ac.kr}

\author{Ig-Jae Kim}
\affiliation{%
  \institution{Korea Institute of Science and Technology}
  \city{Seoul}
  \country{Korea}
}
\email{drjay@kist.re.kr}

\author{Dongwook Yoon}
\affiliation{%
  \institution{University of British Columbia}
  \city{Vancouver}
  \country{Canada}
}
\email{yoon@cs.ubc.ca}

\renewcommand{\shortauthors}{Liu et al.}

\begin{abstract}
    \emph{\textbf{ABSTRACT}} Social media clones are AI-powered social delegates of ourselves created using our personal data. As our identities and online personas intertwine, these technologies have the potential to greatly enhance our social media experience. If mismanaged however, these clones may also pose new risks to our social reputation and online relationships. To set the foundation for a productive and responsible integration, we set out to understand how social media clones will impact our online behavior and interactions. We conducted a series of semi-structured interviews introducing eight speculative clone concepts to 32 social media users through a design workbook. Applying existing work in AI-mediated communication in the context of social media, we found that although clones can offer convenience and comfort, they can also threaten the user’s authenticity and increase skepticism within the online community. As a result users tend to behave more like their clones to mitigate discrepancies and interaction breakdowns. These findings are discussed through the lens of past literature in identity and impression management to highlight challenges in the adoption of social media clones by the general public, and propose design considerations for their successful integration into social media platforms.
\end{abstract}


\maketitle

\fancyfoot{} 
\renewcommand\footnotetextcopyrightpermission[1]{} 

\section{Introduction}
The idea of leveraging generative AI and personal data to represent one’s self, or AI clones, has become a reality within social media. In May 2023, LinkedIn introduced their new AI-assisted messaging feature \cite{srinivasan_helping_2023}, enabling recruiters to create personalized reach-out messages to potential candidates with a click of a button. Meta then took it a step further in September that same year by unveiling a series of interactive AIs to their social media platforms \cite{meta_introducing_2023}; taking on celebrity personas like Tom Brady and Snoop Dogg, these AIs were met with a great deal of curiosity and criticism due to their replication of real life individuals. Meta has decided since then to discontinue this service, instead focusing on AI tools that let users create their own chatbots \cite{peters_meta_2024}. The early integrations of this technology, though varying in success, underscore the need for a deeper understanding of AI clones and their potential to shape our social media experiences, both positively and negatively.

Concerns surrounding clones and the challenges they present to users have been explored by past studies. Lee et al. introduced issues such as doppelgänger-phobia and identity fragmentation \cite{lee_speculating_2023}, demonstrating the clones’ potential to exploit our persona and threaten our self perception. Morris et al. explained how clones can damage the user’s reputation by creating negative perceptions of them, as well as compromise security and privacy through malicious use \cite{morris_generative_2024}. These studies emphasize the importance of designing responsible clones in order to mitigate their risk to our security, self identity, and personal relationships.

The application of AI clones holds particular significance within social media, as the clones by their very nature challenge and reshape our identities—an element at the core of social interactions \cite{tajfel_social_1974}. Despite this, social media has seen a surge in both platform- and user-generated clones \cite{hawkins_how_2023, shepherd_digital_2024, tolentino_snapchat_2023, verma_ai_2023}, making social media an accessible and direct way for the general public to engage with such technologies. The use of these clones may challenge the delicate balance of user- and AI agency on social media, which requires nuanced negotiation to operate effectively \cite{kang_ai_2022}. For many of us, these social platforms have become a valuable source of emotional connection \cite{dhar_understanding_2023, ozok_why_2009}, and any issues that arise can significantly impact our social reputations and relationships \cite{baccarella_social_2018}. Expanding on previous work to unveil the unique effects clones would have within social media will be crucial for their successful integration.

As a result, this study sets out to answer the research question, \textit{What are the benefits and risks of AI clones in social media, along with their ramifications on its users?} Our exploration of AI clones is framed through the lens of past literature in the following domain:

1) Social Media Practices \cite{valkenburg_understanding_2017}: How will AI clones be adopted into peoples’ online presentations and what are the value and risks they create for user agency, self-identity and impression management? The communication theory of identity describes the multilayered nature of identity \cite{braithwaite_communication_2021}, and how inter-layer tension can negatively impact communication \cite{jung_identity_2008, wadsworth_role_2008}. Impression management has also been shown to be a complex process on social media due to the challenges of balancing authenticity with the ideal presentation of oneself \cite{marwick_i_2011, zhao_many_2013, van_dijck_you_2013}. Understanding how users adapt their social media practices to the introduction of clones may provide insight into how we can preserve the agency and identity of social media users.

2) AI-mediated Communication \cite{hancock_ai-mediated_2020}: Within social interactions involving clones, what are factors that determine impression transfer, and their effects on relationship formation and maintenance? Explicit AI-mediated communication has shown both potential for maintaining relationships \cite{hohenstein_ai_2020} as well as undermining interpersonal perceptions \cite{mieczkowski_ai-mediated_2021}. The possibility of impression transfer from AI to humans has also been observed \cite{li_impression_2023}, along with an increase in mistrust, known as the "Replicant Effect," when the use of AI is not disclosed \cite{jakesch_ai-mediated_2019}. These effects may be exacerbated in the case of clones, as personal identity is further manipulated and communication may become entirely with the clone itself. All these factors may have a profound impact on our long term relationship development online, which often values trustfulness and equality \cite{emerson_social_1976}.

To address the speculative nature of AI clones in the social media context, we developed a design workbook featuring eight distinct concepts of clone use in social media. The design workbook was presented to 32 active social media (i.e., Facebook, X, LinkedIn) users in a series of interviews. Through these interviews, we captured their responses to a range of scenarios involving clones. Results from the study revealed both the value clones can bring to productivity and comfort, the risks they pose to the authenticity of social media, and the consequences of those risks on the Target’s impression and relationships.

This study contributes empirical findings on the benefits and risks of clones on our social reputation and relationships, alongside the impact these clones can have on people’s impression management practices and impression transfer mechanisms. Furthermore, it extends relevant theories by empirically demonstrating the nuanced ways impression management, identity negotiation, and relationship dynamics operate when mediated by AI clones in social media.

\section{Related Work}

This work builds on prior research concerning definitions of AI-based human representations, the dynamics of user versus AI agency, theories of identity and impression management, and the impact of AI on trust in social interactions.

\subsection{Defining Social Media Clones}
AI-based digital representation of humans has become commonplace on social media \cite{karnouskos_artificial_2020, westerlund_emergence_2019}. Technologies like deepfakes that uses AI to mimic the voice and visual likeness of a person \cite{amezaga_availability_2022} have caused a fair share of controversy in the online community due to their frequent misuse \cite{mustak_deepfakes_2023, public-private_analysis_exchange_program_increasing_2021}. Going beyond the physical aspects of an individual, AI also has the potential to learn and predict our behavior. Truby and Brown \cite{truby_human_2021} expanded on this concept in the context of marketing and defined what they call a “Human digital thought clone”, where each individual consumer will have a “digital twin” that is created from all the available data of that specific person. They referred to this as the “Holy Grail” for advertisers as it is akin to having access to each consumer’s thoughts, giving companies even greater power to predict and influence consumer decisions.

Although cloning consumer behavior has become a prominent use case due to its commercial value, it is far from the only application that this technology has to offer. Saddik \cite{el_saddik_digital_2018} described another concept of a "digital twin", which can improve an individual's quality of life and enhance well-being by modeling their health and predict illness. McIlroy-Young et al. \cite{mcilroy-young_mimetic_2022} proposed AI models of individuals that are designed for interaction. These researchers coined the term “Mimetic Models”, which describes an algorithm trained with data from an individual in a specific domain, with the goal of simulating their behavior in new situations within that domain. Lee et al. \cite{lee_speculating_2023} had a similar definition, where AI clones are described as interactive agents that represent some aspects of a specific person, built on the personal data of that person.

Finally, we shall refer to AI clones that exist within a social media platform as \textit{Social Media Clones}, defined by the following representative features:

(i) \textit{Each social media clone is built on the data of a Target using artificial intelligence.}

(ii) \textit{Social media clones are generative, responding to external input as the Target would.}

(iii) \textit{Social media clones are interactive through the affordances of the platform they reside in.}

\noindent This definition closely aligns with our design concepts, adapted from a combination of the definitions proposed by prior work \cite{mcilroy-young_mimetic_2022, lee_speculating_2023}, with some modifications to ground it in the domain of social media. What makes these clones unique is their integration with social media affordances \cite{burgess_affordances_2018}. This includes both high level affordances like self-expression and persistence \cite{papacharissi_social_2010, treem_social_2012}, as well as feature specific affordances like the ability to direct message people or publish posts on a timeline \cite{sundar_social_2015}.

\subsection{User Agency vs. AI Agency in Social Media}
AI is a prominent component of many modern social media platforms. When used behind the scenes they can enhance platform affordances such as personalization in content consumption \cite{sadiku_artificial_2021, singh_implications_2023} and online networking \cite{saheb_convergence_2024}. They can also be used directly by the users as a tool for content creation and social engagement \cite{srinivasan_helping_2023, bilhete_ai_2024}. As previously mentioned, Companies like Meta have begun experimenting with AI chatbots for users to talk to. Similarly, Snapchat came up with a feature they call “My AI”, which is a personal chatbot that is unique to each individual, and can be personalized based on the user data it receives \cite{snap_inc_what_2023}. Numerous third party tools have also emerged that help users improve their online image through services like writing their profile bios or generating their profile pictures \cite{bilhete_ai_2024}.

However, this increasing involvement of AI in social media may threaten the agency of its users \cite{kang_feeling_2020}, an important aspect for those who value greater autonomy \cite{bol_customization_2019}. Kang and Lou explained how negotiating user agency and AI agency plays a critical role in social media engagement \cite{kang_ai_2022}. Their research highlights how the two agencies can find synergy through mutual augmentation, where the user strikes a balance between exerting their own agency versus allowing the AI to provide a more convenient or enjoyable social media experience. This idea is particularly relevant with AI clones, as users must find the appropriate amount of agency they are willing to give up for the proper operation of their clones.

\subsection{Identity and Impression Management} \label{identity_and_impression}
Many theories of identity have emerged in psychology and sociology. Our study will primarily focus on Hecht and Phillips’ communication theory of identity (CTI) \cite{braithwaite_communication_2021}, which does not view identity as a static concept, but rather as the summation of 4 interdependent layers: personal, relational, enacted, and communal. An identity gap may occur when there are inconsistencies between two or more identity layers, prompting us to correct one of the layers through our actions or thought process in order to avoid negative communication or relational outcomes \cite{jung_identity_2008, wadsworth_role_2008}. Adopting a social identity perspective allows us to rationalize the self-presentation and intra-group interactions of AI clone users \cite{seering_applications_2018}. Through the lens of CTI, clones can be viewed as an extension of the Target’s enacted identity. This allows us to understand how behavioral mismatches between the clone and Target can influence the Target behavior, and by extension their identity as a whole.

The idea of our identity being something malleable and expressed through our communication also plays a role in how we present ourselves to others. Goffman called this process impression management \cite{goffman_presentation_1990}, which is how we influence other people's perception of us through social interactions. This is especially prevalent on social media platforms, where we have greater control over our self-presentation \cite{walther_computer-mediated_1996}, and the ability to broadcast to our entire social network \cite{marwick_i_2011, zhao_many_2013, van_dijck_you_2013}. In the context of AI-mediated communication, impressions have been found to transfer from the AI onto the user \cite{li_impression_2023}. To avoid being perceived negatively, Endacott and Leonardi \cite{endacott_artificial_2022} discovered how users managed their impressions in AI-MC through: 1. managing how their Interactors interpreted their use of clones (interpretation); 2. managing the relationship between Interactor and AI (diplomacy); and 3. monitored how they communicated with the AI when visible to Interactors (staging politeness). In turn, Interactors had three processes to form impressions of the Target: 1. AI reinforcing existing Target impressions (confirmation); 2. impression of the AI shaping the impression of the Target (transference); 3. viewing the impression of AI separate from the Target (compartmentalization). These processes provide valuable insight into how clones may affect the impression management and transference of Targets and help contextualize participant behavior within our study.

\subsection{The Effects of AI on Trust in Social Interactions}
The lack of trust in AI systems can have a negative impact on social interactions \cite{hohenstein_artificial_2023}. Previous research observed what Jakesch et al. dubbed the "replicant effect", where people had higher mistrust of Airbnb profiles that they perceived to be AI generated \cite{jakesch_ai-mediated_2019}. Liu et al. found a similar effect where participants lowered their trust in email writers when it was revealed to them that an AI was involved in the writing process \cite{liu_will_2022}. To increase trust in AI, studies have shown that large language models can explain their decisions to give users more context \cite{huang_can_2023}, creating a human-in-the-loop system that promotes transparency \cite{li_exploring_2024}. We can also improve confidence in these systems by increasing the user's data agency, giving them the ability to take actions on how their data is being collected and what it will be used for \cite{ajmani_data_2024, pybus_hacking_2015}.

Although our study focuses on text-based AI clones, previous work have looked at using AI agents to represent people in social interactions in the form of generative speech \cite{hwang_whose_2024}. Hwang et al. suggested that the adoption of AI agents may negatively impact human communication, undermine the value of social interaction, and threaten the user’s ability to control their image. The authors further suggests the need for “user-defined red lines” if the technology are to be applied in social settings.

Despite these mistrusts, Kreps et al. describes what they called the "Trust Paradox" \cite{kreps_exploring_2023}, where people’s willingness to use AI-enabled technologies is disproportionally greater than the level of trust they have in the technologies’ capabilities. People still view AI as a promising tool for the future \cite{gerlich_perceptions_2023}. One such example is their ability to serve as a “moral crumple zone” during unsuccessful conversations, where they were attributed some of the responsibility from the interaction partner \cite{hohenstein_ai_2020}. This demonstrates the positive effects AI can have on interpersonal interaction and relationships, and provides encouragement for further development within this domain.

\section{Method}
This study deploys a series of participant interviews with the aid of a design workbook \cite{wyche_benefits_2021, gaver_making_2011}---a set of design documents containing core concepts of social media clones in various usage scenarios. We will be using the following terms defined by McIlroy-Young et al. \cite{mcilroy-young_mimetic_2022} to refer to their corresponding user types in our study. 
\begin{itemize}
    \item Targets: The users who the clone is representing and operated by
    \item Interactors: The users who interacts with the clone, typically within the Target’s social network
\end{itemize}

\begin{figure}[h]
  \centering
  \includegraphics[width=\linewidth]{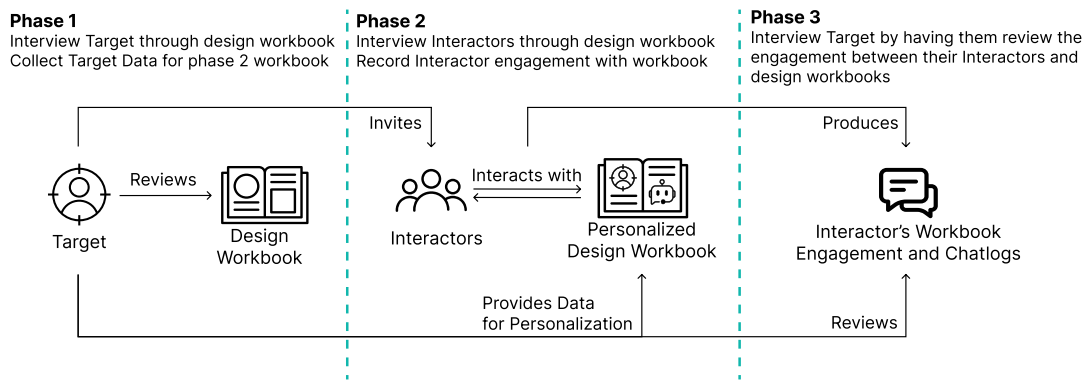}
  \caption{Overview of the 3 study phases designed to iteratively develop the design workbook and illicit Target and Interactor reactions}
  \Description{Figure showing the activities across the 3 interview phases of the study. Phase 1 has Targets review the design workbook, invite Interactors for phase 2, and provide data for workbook personalization. Phase 2 has Interactors engage with the personalized workbook, from which engagement data is collected for phase 3. Phase 3 invites Target back to review their Interactor's engagement with the workbook.}
  \label{fig:study-overview}
\end{figure}

\subsection{Design Workbook}
Design workbooks enable us to generate speculative design concepts \cite{yelavich_speculative_2015} of clone use in social media, introducing the novel idea to our participants through real-world examples and intuitive visuals. Besides being a cost-effective solution to the technical challenge of developing fully functional clones for each participant, the creativity and freedom fostered by the workbook also facilitates in depth discussion and exploration within this new domain.

When determining the context for clone integration, we focused on four fundamental elements proposed by Bayer et al. \cite{bayer_social_2020} to be common across most social media platforms: 1. Profile, 2. Network, 3. Stream, and 4. Message. This approach maintains the theoretical relevance of our study despite the ever-changing social media landscape and the diverse features they offer \cite{aichner_twenty-five_2021, zhang_form-_2024}.

To develop the design workbook, two ideation workshops, each lasting one hour, were conducted to generate 12 unique initial concepts. The first workshop involved three members of the research team, where we formed initial ideas for the use of clones in social media. The second session added six HCI researchers, including two PhD students, two Master’s students, and two undergraduate researchers. With the lead investigator as the facilitator, the group brainstormed concepts that are scaffolded by the initial ideas from the first workshop but to expand and refine them.

\begin{figure}
\centering
\begin{subfigure}{.5\linewidth}
  \centering
  \includegraphics[width=\linewidth]{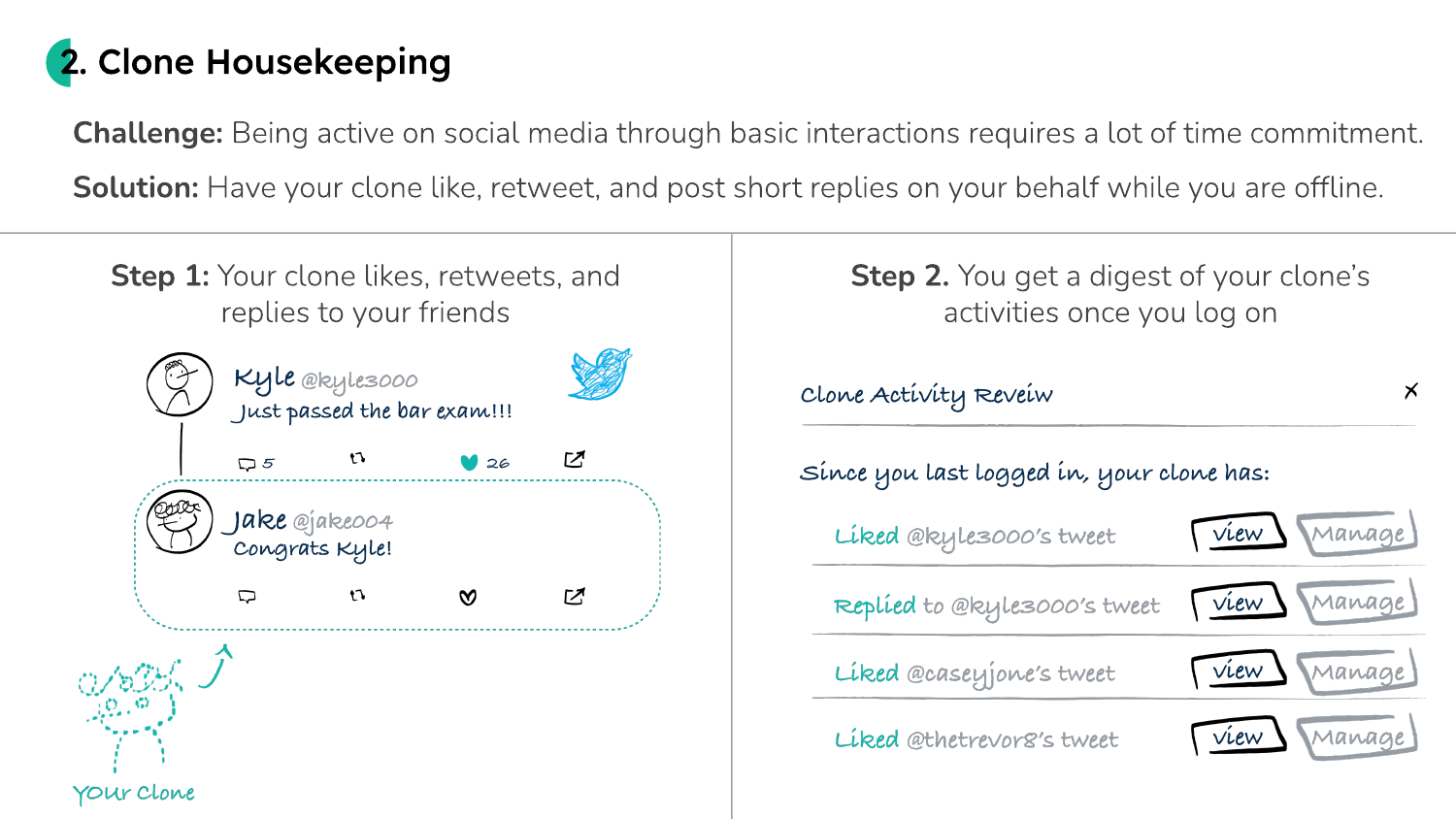}
  \caption{Design workbook page for the concept \textit{Clone Housekeeping}}
  \Description{Workbook page begins with the name of the concept "Clone Housekeeping", followed by a sentence describing the challenge as "Being active on social media through basic interactions requires a lot of time commitment." After that a sentence describing the solution involving AI clones: "Have your clone like, retweet, and post short replies on your behalf while you are offline." Under the text description is an illustration of this concept. Step one shows a sketch of your clone replying to one of your friend's tweets on your behalf. Step two shows a summary list of all the activities your clone conducted while you were offline, giving you the option to view and manage them.}
  \label{fig:sub1}
\end{subfigure}%
\begin{subfigure}{.5\linewidth}
  \centering
  \includegraphics[width=\linewidth]{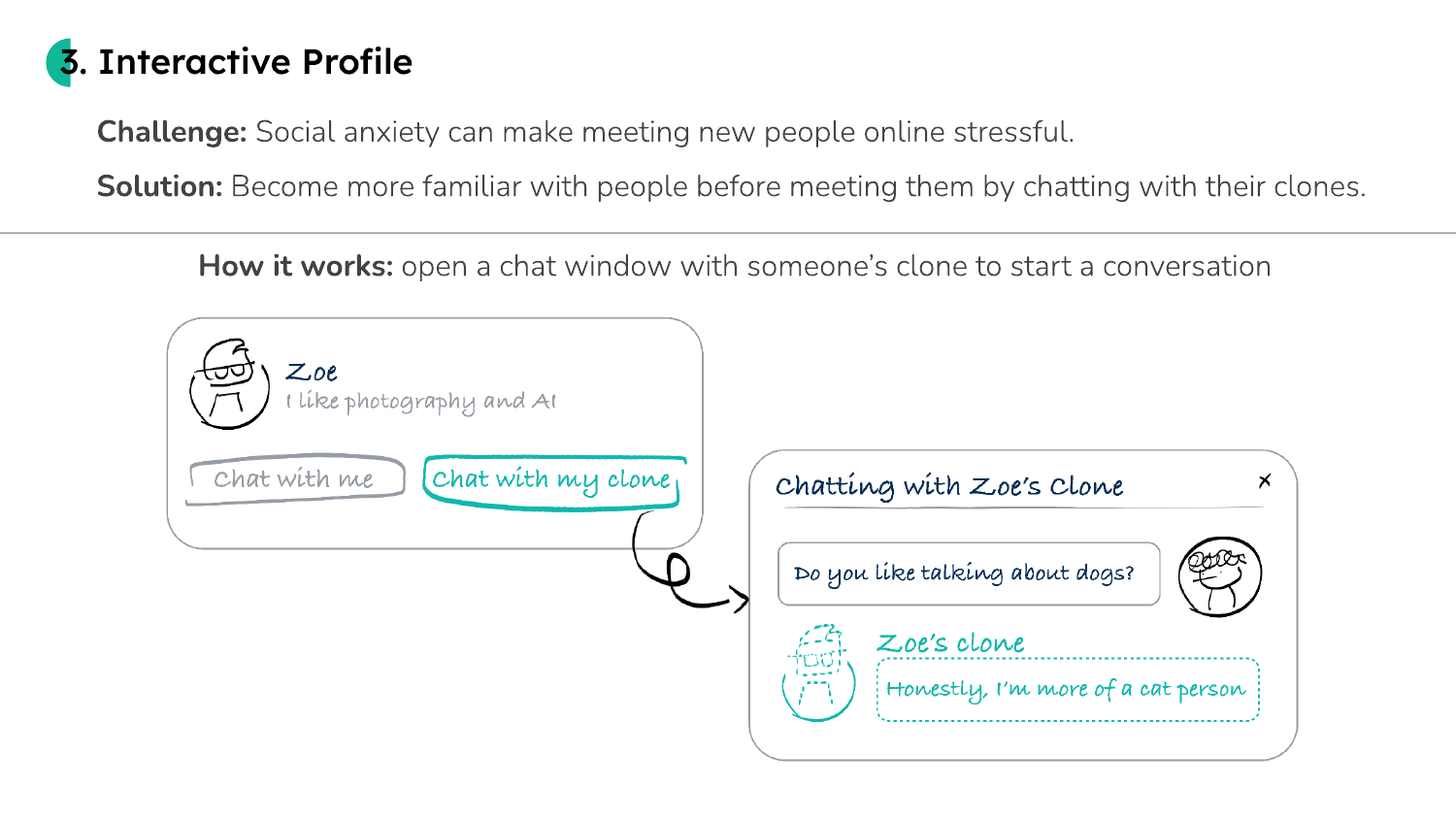}
  \caption{Design workbook page for the concept \textit{Interactive Profile}}
  \Description{Workbook page begins with the name of the concept "Interactive Profile", followed by a sentence describing the challenge as "Social anxiety can make meeting new people online stressful." After that a sentence describing the solution involving AI clones: "Become more familiar with people before meeting them by chatting with their clones." Under the text description is an illustration of this concept. A sketch of a Target's profile is shown with a button named "Chat with my clone". An arrow from the button points to a chat window that would pop up upon clicking the button, which holds a conversation between the Target's clone and the Interacrtor.}
  \label{fig:sub2}
\end{subfigure}
\caption{Example pages of the design workbook used in phase 1 interviews}
\label{fig:workbook}
\end{figure}

The research team then ranked the concepts based on their potential to elicit rich participant responses. This is determined by categorizing the attributes, behavior, and value of each concept based on prior literature in AI-mediated communication and social computing (Appendix \ref{apx:concept-table}). Such aspects include the social media element the concept operates in \cite{bayer_social_2020}, the concept's degree of AI autonomy \cite{hancock_ai-mediated_2020, kim_one_2023}, the value that the concept provides \cite{hu_effect_2015, dhar_understanding_2023}, and the optimal relationship the concept targets \cite{planalp_friends_1992, funder_friends_1988, kairam_talking_2012}. A design workbook consisting of eight final concepts of social media clones was created (Table \ref{tab:concepts}) where concepts, collectively, cover diverse types of attributes, behaviors, and value. Each concept features a written description of a clone use case, accompanied by an illustration of how the concept would work to aid with participant comprehension (Figure \ref{fig:workbook}).

\emph{Social Media Clone Chatbot.}
A clone chatbot of the Target was added to the phase 2 workbook as a variant of the \textit{Interactive Profile} concept to further concretise participant perception of social media clones. These chatbots were created with prompt engineering, leveraging the Target’s provided conversation data as input, and implemented using GPT (API version: gpt-4, accessed between January-May 2024). Interactors, recruited through a Target participant, were able to converse with the clone of their Target through a social media messaging web application. This lets them experience what engaging with a clone on social media would feel like. Targets are also able to get a sense of what having their clones interact with others would feel like when they review the chat logs in phase 3. The prompt for creating the clone is in Appendix \ref{apx:prompt}.

\begin{table}
  \small
  \begin{tabular}{lp{0.8\linewidth}}
    \toprule
    \textbf{Concept Name} & \textbf{Concept Description}\\
    \midrule
    1.Interactive Profile & The clone analyzes your past interactions and profile data to mimic your online persona, it can then have conversations with your social network\\ \hline
    2. Cross Media Posting & The clone synchronizes all your social media postings by creating a version of the original post for each platform according to their appropriate formats\\ \hline
    3. Clone Housekeeping & The clone helps you remain active online by engaging and interacting with your friend’s online activity through likes/reactions and comments/replies\\ \hline
    4. Personalized Reachout & The clone initiates a personal introduction conversation to new contacts to scale the expansion of your social network\\ \hline
    5. Undercover Rejection & The clone helps you avoid uncomfortable social situations by generating and delivering messages to turn down invitations.\\ \hline
    \vtop{\hbox{\strut 6. Post-completion for}\hbox{\strut Trending Topics}} & The clone creates a new social media post based on a given prompt and your previous post activities for you to engage with trending topics online\\ \hline
    7. Mood Modifier & The clone modifies your replies in an online conversation in real time to achieve certain effect such as being more polite or professional\\ \hline
    8. Group-convo Simulation & The clone can partake in a discussion with the clone of others on a topic of your choosing, such as seeing a conversation between the clone of two scientists about their domain\\
  \bottomrule
\end{tabular}
\caption{Summary of the eight social media clone concepts from the design workbook}
\label{tab:concepts}
\end{table}

\subsection{Social Media Platforms}
Studies have found that different social media are used in functionally different ways and each may have a unique relation to friendship closeness \cite{pouwels_social_2021}. To assess clones in diverse social media contexts, we selected three representative platforms: Facebook for close relationships and casual interactions, X (formerly, Twitter) for a mix of casual and professional interactions, and LinkedIn for professional relationships. These platforms cover a broad range of social media elements while minimizing overlap in interaction types and relationship dynamics. They were also chosen for their high number of monthly active users, ensuring broad applicability of our findings. Other platforms like WhatsApp, WeChat, Snapchat, YouTube, TikTok, and Instagram were considered but excluded due to their narrow focus on specific elements, or prioritizing content creation over social interactions.

\subsection{Participants}
We recruited participants through various social media platforms, including Facebook, X, and LinkedIn, ensuring a broad reach across different types of users. Among those who volunteered, we employed purposive sampling to ensure diversity and equal representation across the platforms. The recruitment posts included a brief description of the study, and a link to a survey for interested participants. 32 total participants were recruited across all phases, with which 38 interviews were conducted. See the breakdown of participants per each phase in Table \ref{tab:participants} and the demographic breakdown in Table \ref{tab:demographic}.

\begin{table}[]
\small
  \begin{tabular}{lll}
    \hline
    \textbf{\begin{tabular}[c]{@{}l@{}}Phase and\\ Participants\end{tabular}} & \textbf{\begin{tabular}[c]{@{}l@{}}Platform\\ Breakdown\end{tabular}} & \textbf{\begin{tabular}[c]{@{}l@{}}Recruitment\\ Details\end{tabular}} \\ \hline
    \begin{tabular}[c]{@{}l@{}}Phase 1\\ 14 Target Participants\end{tabular} & \begin{tabular}[c]{@{}l@{}}4 Facebook users\\ 6 X users\\ 4 LinkedIn users\end{tabular} & \begin{tabular}[c]{@{}l@{}}Six willing Targets (two per platform) from this phase\\ were randomly selected to participate in phases 2 and 3\end{tabular} \\ \hline
    \begin{tabular}[c]{@{}l@{}}Phase 2\\ 18 Interactor Participants\end{tabular} & \begin{tabular}[c]{@{}l@{}}6 Facebook users\\ 6 X users\\ 6 LinkedIn users\end{tabular} & \begin{tabular}[c]{@{}l@{}}The six Targets from phase 1 each recruited three of their friends\\ on their respective platforms to participate as the 18 Interactors in phase 2\end{tabular} \\ \hline
    \begin{tabular}[c]{@{}l@{}}Phase 3\\ 6 Target Participants\end{tabular} & \begin{tabular}[c]{@{}l@{}}2 Facebook users\\ 2 X users\\ 2 LinkedIn users\end{tabular} & \begin{tabular}[c]{@{}l@{}}The six Targets were invited back to review how their Interactors'\\ responded and interacted with the personalized workbooks\end{tabular} \\ \hline
  \end{tabular}
  \caption{Breakdown of participant recruitment across study phases and social media platforms}
  \label{tab:participants}
\end{table}

\begin{table}
\scriptsize
\centering
\begin{tabular}{lllllll}
\hline
\textbf{\begin{tabular}[c]{@{}l@{}}Target (T) /\\ Interactor (I)\end{tabular}} & \textbf{Gender} & \textbf{Age} & \textbf{Ethnicity} & \textbf{AI Familiarity} & \textbf{\begin{tabular}[c]{@{}l@{}}Years of\\ Social Media Use\end{tabular}} & \textbf{\begin{tabular}[c]{@{}l@{}}Frequency of\\ Social Media Use\end{tabular}} \\ \hline
T1 & Man & 25-34 & Black & Expert & 5-10 years & Daily \\
T2 & Man & 25-34 & Black & Expert & 5-10 years & Daily \\
T3 & Woman & 25-34 & Asian & Beginner & 10+ years & 4-6 times a week \\
T4 & Man & 19-24 & Black & Intermediate & 1-5 years & Daily \\
T5 & Man & 25-34 & Black & Expert & 5-10 years & Daily \\
T6 & Woman & 25-34 & Asian & No experience & 1-5 years & 4-6 times a week \\
T7 & Man & 19-24 & Asian & Beginner & 1-5 years & 2-3 times a week \\
T8 & Man & 19-24 & Asian & Intermediate & 1-5 years & 4-6 times a week \\
T9 & Man & 25-34 & Asian & Beginner & Less than 6 months & 2-3 times a month \\
T10 & Man & 25-34 & Black & Expert & 10+ years & Daily \\
T11 & Man & 25-34 & Black & Expert & 10+ years & Daily \\
T12 & Man & 25-34 & Black & Expert & 10+ years & Daily \\
T13 & Woman & 25-34 & Asian & Beginner & 10+ years & Daily \\
T14 & Woman & 19-24 & Asian & Beginner & 5-10 years & Daily \\
I1 & Woman & 25-34 & Asian & Beginner & 10+ years & Daily \\
I2 & Man & 25-34 & Asian & Beginner & 10+ years & Daily \\
I3 & Man & 25-34 & Asian & Beginner & 10+ years & Daily \\
I4 & Woman & 19-24 & Asian & Intermediate & 10+ years & Daily \\
I5 & Woman & 19-24 & Asian & Beginner & 5-10 years & Daily \\
I6 & Man & 19-24 & Asian & Intermediate & 5-10 years & Daily \\
I7 & Man & 19-24 & Asian & Beginner & 1-5 years & Daily \\
I8 & Man & 25-34 & Asian & Expert & 10+ years & Daily \\
I9 & Woman & 25-34 & Asian & Beginner & 10+ years & Daily \\
I10 & Man & 25-34 & Asian & Beginner & 10+ years & Daily \\
I11 & Man & 25-34 & Asian,European & Beginner & 10+ years & Daily \\
I12 & Woman & 25-34 & Asian & Beginner & 10+ years & Daily \\
I13 & Woman & 19-24 & Asian & Intermediate & 5-10 years & Daily \\
I14 & Non-binary & 19-24 & Asian & Beginner & 5-10 years & Daily \\
I15 & Woman & 19-24 & Asian & Beginner & 10+ years & Daily \\
I16 & Man & 25-34 & European & Expert & 10+ years & 4-6 times a week \\
I17 & Woman & 25-34 & Asian & Beginner & 10+ years & Daily \\
I18 & Prefer not to say & 25-34 & Asian & Beginner & 10+ years & Daily \\
\hline
\end{tabular}
\caption{\label{tab:demographic} Summary of participant demographics}
\end{table}

\subsection{Data Collection}
Data collection consisted of three sets of interviews: one hour interviews with the 14 Targets, one hour interviews with the 18 Interactors, and 45 minute interviews with the six  Targets who recruited their friends for phase 2. All interviews were conducted through Zoom with video/audio recording. Participants in phases 1 and 2 received CAD \$40 for completing the interview, while participants in phase 3 received an additional CAD \$30.

\textit{Phase 1 - Target Interviews} began with an introduction and definition of social media clones, followed by a walkthrough of the design workbook where the
interviewer explains the concepts to the participants. Participants provided their first impressions after reviewing each concept in free-form dialogue. The explanation and discussion of one concept usually took 3-4 minutes to complete, and
reviewing all eight concepts of the design workbook took 25-30 minutes in total per
interview. Participants were asked additional questions at the end of the walkthrough. These questions touch on their perceptions of clones, and the effects that the clone can have on their impression management practices. Interview questions for all 3 phases can be viewed in Appendix \ref{apx:questions}. At the end of the session, six willing participants were sampled at random to participate in subsequent phases. We requested each selected Target to invite three of their social media friends or followers for participation in phase 2, and to provide 100 messages from their social media account to serve as training data for the development of their clone chatbot.

\textit{Phase 2 - Interactor Interviews}  had six out of the eight concepts developed further for phase 2. \textit{Mood modifier} was removed due to its similarity to existing tools such Gmail’s smart replies, decreasing likelihood of novel findings; \textit{Group Convo Simulation} was removed as it had the lowest level of participant engagement from phase 1. For each set of Interactors, a personalized variant of the workbook was created, substituting the low-fidelity sketches with higher-fidelity social media mockups of the Target. This provided a more immersive demonstration of the concepts and elicited more concrete feedback. The interview also included an interactive component by having participants partake in a five minute conversation with their Target’s clone chatbot. The interviewer informed the participants that they were engaging with an AI clone for this activity, to reflect the intended disclosure of the \textit{Interactive Profile} concept. The questions in this phase focused on the Interactor's perspective, exploring themes of impression transfer, and how they interact with the clones.

\textit{Phase 3 - Target Follow-up Interviews} invited back the six Target participants from phase 1. They are presented with the chatlog between their Interactors and their chatbot, along with any noteworthy comments from the Interactors regarding the design workbook. Upon reviewing the Interactor engagement with the workbook, they are asked a series of questions on their experience as the owner of the clone. The session concludes with any final comments and questions.

\subsection{Data Analysis}

Qualitative Data Analysis was conducted through reflexive thematic analysis \cite{braun_using_2006} of the 38 interview transcripts. This aligns with the exploratory and iterative nature of our method, enabling us to adapt our analysis to data across different study phases and cover a diverse range of potential findings. Analysis was conducted in parallel to the participant interviews, as to allow for addressing any identified gap in the data by updating the interview questions. This was especially useful for findings that seemed promising but required more clarification and insights, where the updated interviews were able to provide.

A preliminary coding of a subset of transcripts was first conducted by the lead researcher. After verifying the quality and depth of the codes with the rest of the team, the lead researcher then completed analyzing the remaining transcripts, developing on the existing codes. 59 unique codes were identified after combining similar sentiments. Some examples include “Disclosure of clone use” and “Inappropriate context or tone”. From these codes 10 overarching themes emerged both inductively and deductively. Inductive themes were directly extracted from the code, such as “Causes of breakdowns in clone interaction”. Deductive themes were created through extending the identified codes to findings from prior work, including “Impression transfer from clone to Target”. Thematic saturation was reached once the existing findings received ample clarification and support, and no significant new themes were formed in the last batch of interviews. Weekly meetings were held throughout the study with the entire research team to organize and refine the themes. Through multiple iterations, we identified 9 sublevel claims to be novel and insightful to report in this paper. This includes 3 subthemes about the benefits and risks of clone use, 3 subthemes on the Target’s impression management practices, and 3 subthemes on the impression transfer mechanisms and Interactor response.

\section{Findings}

We found that participants saw the potential for social media clones to reduce effort for relationship maintenance and content generation, while also being able to delegate undesired tasks and increase their online availability to others. In terms of anticipated risks, authenticity was the most prominent concern, as participants felt clones would jeopardize real social connections, and create a level of skepticism in the online community.

When asked about changes to their online impression management practices, Targets feared clones may exacerbate context collapse, in which case they would take responsibility and opt to resolve the conflict directly with Interactors. Targets also believed they would adopt positive behaviors from their clones as a means of self improvement, similar to reflecting on old messages and actions from their past. From the Interactor’s side, they believed that the transfer of impressions from clone to Target is increased based on how positive their perception of clones are and their unfamiliarity with the Target. When this impression is inaccurate, a false impression may be formed leading to loss of trust and interaction breakdowns. We found that Interactors are less likely to address breakdowns than Targets, only choosing to do so in close relationships with a high severity breakdowns. An overview of the findings are summarized in Table \ref{Tab:Findings}.

\begin{table}
\centering
\renewcommand{\arraystretch}{1.5}
\footnotesize
\begin{tabular}{p{1.30in}p{4.40in}}
\hline
\textbf{Theme} & \textbf{Findings} \\
\hline
\begin{tabular}[c]{p{1.30in}}\ref{trade-offs} The Trade-Offs: Convenience and Comfort at the Cost of User Agency and Authentic Social Interaction \end{tabular}& \begin{tabular}[c]{@{}l@{}} \ref{effort_imbalance} Automating relationships reduces effort from Targets but not Interactors, breaking reciprocity expectations \\ \ref{jeopardize_authenticity} Generated posts ease up Target's burden but also jeopardize authenticity perceived by Interators \\ 
\ref{one-sided_relationships} Clones increase Target availability but risks one-sided relationships and privacy \end{tabular} \\
\hline
\begin{tabular}[c]{p{1.30in}} \ref{practices_challenges} Challenges to Target Identity and Impression Management Practices \end{tabular} & \begin{tabular}[c]{@{}l@{}} \ref{adopt_behavior} Adopting clone behavior facilitates impression management but challenges personal identity \\ \ref{context_collapse} Clones can exacerbate context collapse in social media and extend it to one-on-one interactions \\ \ref{shoulder_blame} Targets shoulder the blame for breakdowns and engage directly with Interactors to resolve them \end{tabular} \\
\hline
\begin{tabular}[c]{p{1.30in}} \ref{mechanisms} Impression Transfer Mechanisms, Breakdowns, and Interactor Response \end{tabular} & \begin{tabular}[c]{@{}l@{}} \ref{impression_transfer} Positive perceptions of the clones and unfamiliarity with the Target promotes impression transfer \\ \ref{false_impressions} Mismatch between Target and Clone can lower Interactor trust and create false impressions \\ \ref{avoid_addressing_breakdowns} Interactors tend to avoid addressing breakdowns unless the severity is high and the relationship is close \end{tabular} \\
\hline
\end{tabular}
\caption{Summary of findings}
\label{Tab:Findings}
\end{table}

\subsection{The Trade-Offs: Convenience and Comfort at the Cost of User Agency and Authentic Social Interaction} \label{trade-offs}
Participants recognized that clones have the potential to enhance social media experiences by providing convenience and comfort, such as task delegation for Targets or increased social engagement for Interactors. However, they also foresee negative impacts on relationships, user agency, and authenticity if the clones are misused. These risks are often in direct conflict with the value that the concepts create, and navigating these trade-offs is essential for adopting clones into social media. 

\subsubsection{\textit{Effort imbalance: automating relationships reduces effort from Targets but not Interactors, breaking reciprocity expectations.}} \label{effort_imbalance} While Targets can likely benefit from delegating their social tasks to the clone, Interactors expressed concerns about the imbalance of effort. The reduced effort from the Target can make Interactors feel undervalued, potentially straining the relationship.

Targets saw the potential for clones to interact positively with their social network on their behalf. Concepts like \textit{Personalized Reachout} and \textit{Clone Housekeeping} can enable them to make new connections and maintain existing relationships with significantly lower effort compared to the traditional approach. The delegation of these tasks resonated with many participants (T1, T3, T5, T7, T8, T9, T10, T11, T12, T14):
\begin{quote}
    \textit{“I want to support [my friends], but obviously I don't want to be scrolling on social media constantly being like, ‘oh, did anyone post anything?’ So that part is super useful.”} (T11 on Clone Housekeeping)
\end{quote}
\noindent The general sentiment among Targets is that the clones may allow people to stay engaged with their friends’ online activities and gain new connections without having to constantly be online and monitoring their social media. This is ideal for people that are busy, or if they wish to take a break from social media without feeling like they are ignoring their relationships.

Meanwhile, Interactors perceived the act of delegation as a lack of reciprocal efforts for maintaining the social interaction. Several participants (I5, I6, I11, I14, I17, I18) claimed that they would feel displeased if they were to be an Interactor in the aforementioned two concepts, with I5 claiming it can “impact how [they] would even want to continue a relationship with [the Target]”. Social exchange theory \cite{emerson_social_1976} may suggest that Interactors not only value the interaction quality itself, but also the amount of perceived effort that the Target is putting into the relationship. Interactors attribute the lack of effort to them being undervalued or deprioritized by the Target, which can deteriorate the relationship over time despite them receiving more engagement through the use of clones. These responses indicate that human-to-human interaction is still a necessary step in nurturing meaningful relationships, one that cannot be substituted with clones currently.

\subsubsection{\textit{Generated posts ease up Target's burden but also jeopardizes Target agency and authenticity.}} \label{jeopardize_authenticity} Participants highlighted the promise of AI in online content generation, noting its ability to save time for Targets and create engaging material for Interactors. Participants also feared that overproducing AI-generated content could lead to skepticism and distrust within the online community. Targets expressed concerns about appearing less genuine, while Interactors were troubled by the inability to distinguish between AI- and user-generated content.

Leveraging clones for content generation can range from giving sentence level suggestions and modifying tones such as \textit{Mood Modifier}, to creating full messages and posts on the Target’s behalf like in \textit{Cross Media Posting}. Participants (T3, T5, T7, T8, T12, T13, I1, I7) agreed that these features can make social media more “streamlined”, and “reduce the work that is required for humans”, as T13 pointed out. The increase in Target engagement may also benefit Interactors, as they would have more content to perform relationship maintenance with \cite{shklovski_friendship_2015}.

However, many participants felt that the clone’s ability to alter the Target's original sentiments may challenge the Target's agency and make them appear less genuine. 
This dual concern is particularly evident in cases like \textit{Post-completion for Trending Topics} where the clone engages in topics unfamiliar to the Target. As T10 and others (I8, I9, T5, T7, T9) expressed:
\begin{quote}
\textit{"In this day and age, people really care about what you say and do on social media. [...] So I feel like it's quite important to just make sure that things you actually post are authentic and true to your own feelings or opinions."}
\end{quote}
This quote highlights both the agency and authenticity concerns. From an agency perspective, the Target loses their ability to make informed decisions about their social media presence when clones post about unfamiliar topics on their behalf. From an authenticity standpoint, the clone-generated content may not reflect the Target's genuine feelings or opinions, creating a disconnect between their true self and their online persona.

The social media affordances of persistence and broadcast amplify these concerns. When clones post content that misaligns with Targets' views, these posts become permanent records visible to their entire social network. This not only compromises their agency by removing their control over their online narrative but also undermines their authentic self-presentation. Such dual threats can make users hesitant to employ clones for content creation or modification.

These concerns extend beyond individual users to affect the broader social platform ecosystem. Participants I4, I7, I18, T7, T11 among others brought up how the ambiguity of whether something is generated by a clone would make them more wary when engaging with people and posts on social media, with I18 claiming it could “introduce a level of skepticism online”. As suggested by I18’s comment, even when clone use is not present in an interaction, the knowledge that clones exist may be enough to increase distrust in the social media platform.

\subsubsection{\textit{Clones increase Target availability but risks one-sided relationships and privacy.}} \label{one-sided_relationships} Clones can be a useful source of information for Interactors to learn more about the Target. However, some participants realizes the privacy risks, as the lopsided dynamic may give Interactors the impression that their relationship with the Target is closer than it actually is. Furthermore, information gained through the Target's clone could also be misused for manipulation and social engineering.

Participants found that interacting with the clone allowed them to quickly get to know the Target without needing direct engagement. This enhances the existing social media affordance of gaining information and searchability. The clone acted like a personal wiki of the Target, helping Interactors gather basic information and identify common interests. Its attitude and tone also gave Interactors a sense of what conversing with the Target would feel like. I11 praised the clone’s usefulness, noting that it provided timely responses for those trying to learn more about the Target:

\begin{quote}
    \textit{“I would say the clone is really good and engaging because if you are coming from the perspective of somebody who's just trying to find out more about another person by looking at their profile, the responses are actually really fast and relevant.”}
\end{quote}

\noindent The information provided by the clone, combined with its responsiveness, helps Interactors assess whether to pursue a connection with the Target, demonstrating the clones' value in facilitating social connections.

On the other hand, some Targets realized the clones may be taking these affordances too far, and commented on the potential drawbacks of having a tool with so much information about a person be publicly available on social media. The autonomous nature of clones can make it hard for Targets to control what information is disclosed to whom, creating serious privacy concerns. T11 talks about the danger of fostering “stalker tendencies” by learning about the Target without their knowledge. In more extreme cases, clones have the potential to foster parasocial relationships between the Target and Interactor \cite{hoffner_parasocial_2022}. The lack of control in what information is being disclosed can also lead to malicious use cases. Some participants like T13 brought up how the clones can become a tool for manipulation and social engineering \cite{salahdine_social_2019} by “understanding the psyche of the target”. Although the consequences of these scenarios may be dire, they are also easily avoided if the appropriate guardrails are put in place. The majority of participants did not raise this concern of using clones with the assumption that they would be able to control what information the clone will have access to.

\subsection{Challenges to Target Identity and Impression Management Practices} \label{practices_challenges}

Social media clones are a new avenue in which users can communicate and express themselves online. In the view of CTI, this new form of expression can have a direct impact in shaping the Targets identity. Our study reinforced this idea by highlighting the friction Targets experienced when their clone's behavior clashed with their personal, or enacted layer of identity. In addition to this internal conflict, Targets also expressed concerns about the clone's impact on their relational identity, as previous studies have shown that AI has the ability to influence the impression of their users through a process called impression transfer \cite{mieczkowski_ai-mediated_2021, endacott_artificial_2022, li_impression_2023}. Targets believed impression transfer may be even more pronounced due to the synonymy between themselves and their clone. As such, how Targets operate their clones and negotiate agencies between the user and AI becomes a crucial part of their impression management practice on social media. Targets discussed the impression management strategies they would deploy to balance being true to themselves, and bridging the potential impression gap between them and their clone.

\subsubsection{\textit{Adopting clone behavior facilitates impression management but challenges personal identity.}} \label{adopt_behavior} We surfaced two primary motivations for Targets to behave more like their clones. The first is their desire for continuous self-improvement, where they would adopt behaviors from the clone that led to positive interactions. The second and more challenging motivation is the Target's desire to appear consistent online, where they want to maintain the impression of their clones to their Interactors. Although both these processes are in aid to the Target’s impression management practices, participants found that being influenced by their own clones challenged their sense of self-identity and agency.

Many participants thought reflecting on their clone’s interactions would help them identify areas of improvement in their own behavior. Although the Targets were wary of being influenced by their clones rather than the reverse, this motivation was easier for participants to justify as they are resolving the inconsistency within their identity by updating their behavior (enacted identity) to better align with how they perceive themselves (personal identity). T10 talks about this dilemma and how they would rationalize their decision:

\begin{quote}
    \textit{“I do find that it's a bit of a challenge to your own personality, right? But I feel like if you pick the traits that you wanted to work on anyways, and then you just learn how to improve those traits based on your clone. I feel like that's more like personal growth than a really big challenge to our identity."}
\end{quote}

\noindent This participant proposed to only adopt aspects they already wanted to improve on. By doing so, they are able to align their own agency with the agency of their clone. This echos Kang and Lou's idea of mutual augmentation \cite{kang_ai_2022}, whereby the Target learns from their clone's behavior to better themselves, and the clone uses the Target's behavior to increase its accuracy. Through this synergy of agencies the Targets can overcome the threat to their identity and engage in a process of self-improvement.

Participants also expressed their inclination to adopt clone behavior due to their desire to appear consistent online. This effect mirrors the concept of public commitment in social media discussed by Valkenburg \cite{valkenburg_understanding_2017}, where individuals are influenced by their public self-presentation and feel motivated to maintain that representation. If the Interactor has a certain expectation for how the clone behaves, Targets felt the pressure to maintain that impression during their own interactions. In instances where the clone's behavior already aligns with the Target's enacted identity, the clone could act based on the Target, and then reinforce that behavior back to the Target. The issue arises when the clone's behavior does not match with the Target's enacted identity. I4 mentions how they want to “keep the same level of interaction with [the Interactor]” but also question if they are actually themselves or “trying to be like [their] clone to come off more streamlined”. In this scenario, Targets can still behave more like their clone. However, Unlike the previous case, this update in enacted identity will not necessarily align with their personal identity. This mismatch in the Target's identity layers can create a potential identity gap, causing negative emotions in the Target as they choose between appearing consistent online and being true to themselves.

\subsubsection{ \textit{Clones can exacerbate context collapse in social media and extend it to one-on-one interactions.}} \label{context_collapse} One common concern Targets have with deploying clones into their social media is how they may interfere with the Targets' existing impression management process, particularly in curating content for specific audiences. Navigating these different audiences from across their social media platforms is a complicated process that requires a good grasp of the intricacies in each relationship and situation. Participants expressed their doubt that a clone as advanced as it may be, would be able to fully replicate that process. Failure to represent the Target accordingly could lead to context collapse \cite{goffman_presentation_1990} and negatively impact Interactor perceptions of the Target \cite{heider_perceiving_1958}. Our findings discovered the main concern for context collapse is the clone applying an inappropriate context to an interaction or social media platform.

In the case of context collapse in a specific interaction, the clone may be accurately acting like how the Target would in a general sense, but it does not behave accordingly given the specific interaction that it is currently operating within. This is distinct from the situation in the previous section, as there is no conflict between the clone's actions and the Target's enacted identity, but rather the clone now challenges the Target's relational identity. TI4 explained this concern by comparing their interaction between a close friend and an acquaintance:

\begin{quote}
    \textit{“There's a lot of nuances and social cues in terms of what's socially acceptable, even certain things that I might say to a childhood friend in a conversation versus what I might say to a colleague that I just met a week ago. And a clone, maybe they don't have those nuances in the beginning [...].”}
\end{quote}

\noindent In this hypothetical example, the clone is trained on conversations between the Target and their childhood friend. If this clone then talks to a new colleague of the Target, it would continue behaving as though it was talking to a close friend, potentially breaking social norms. Most social media offer users the affordance of segregating their content for different audiences, such as Facebook's friends lists, or even to a specific individual in the form of a direct message. This allows users to switch to the appropriate context for each situation, without having to worry about pleasing different audiences all at once. Though clones may be able to utilize these features, they may not have the situational awareness to take advantage of them, thus reintroducing the problem of context collapse back into the platforms.

Context collapse can go beyond a specific interaction or person. The social media that the clone is deployed in can also determine the behavioral expectations of the clone. One such example participants have thought of was on LinkedIn, where they would generally express themselves in a professional manner, in contrast to their behavior on Facebook or X. As a result, they would also expect their clone to “cater to its audience” as I14 and others (T1, T3, T13) described. The two situations for context collapse poses a big challenge to the Targets’ impression management, and is one of the reasons many Targets prefer direct supervision of the clones' actions to ensure they are operating within the appropriate context.

\subsubsection{\textit{Targets shoulder the blame for breakdowns and engage directly with Interactors to resolve them.}} \label{shoulder_blame} An interaction breakdown may occur when the clone behaves inappropriately, such as in the case of context collapse. When this happens, Targets universally expressed that they would take responsibility and attempt to rectify the situation with the Interactors. This closely resembles the impression management practice of "diplomacy" mentioned by Endacott and Leonardi, where users try to mediate the relationship between their AI and the Interactor, minimizing the damage to their relationships and online reputation.

Majority of the Targets viewed reaching out to the Interactor directly during a breakdown to be the best course of action. Targets believed this allows them to resolve the breakdown, reiterating that the clone was not accurately representing them in the prior interaction. Personally stepping in also ensures that the clone will not further jeopardize the situation. T13 suggested to “take responsibility for making the choice to use the clone” and also “apologize for the opinion expressed”. Other participants like T3, T9, and T12 also highlighted the importance of being honest and transparent during this process, as breakdowns with the clone can negatively impact the Target’s authenticity and truthfulness. By being candid during the conflict resolution, Targets hope to dampen the reputational damages done by their clone.

Targets typically shared the blame of breakdowns between themselves and the clone’s developers. Targets felt the developers should receive partial blame as they are the ones to design and create the clone, and any errors that occur were considered a mistake on their part. The Targets also blamed themselves for choosing to use the technology, and providing the data that resulted in the breakdown.

\begin{quote}
    \textit{“I feel like it'd be more so on the people who developed it or researched how to create it. If I consented to this then it is kind of on me because I'm OK for this to be used in whatever way it's being used”} (T9 on who should be responsible for breakdowns)
\end{quote}

\noindent T9 explained that although the Interactor and clone were the actual participants in the interaction, they were not held responsible for the breakdown. This is distinct from interactions between people, where responsibility is typically shared amongst the people involved. The AI not being held responsible also contrasts with past findings in AI assisted interactions, where they often absorb some of the blame from the user \cite{hohenstein_ai_2020}. However, this finding does align with the idea that Targets view their clones as an extension of their own online representation rather than an external tool, as discussed by Lee et al. \cite{lee_speculating_2023}.

\subsection{Impression Transfer Mechanisms, Breakdowns, and Interactor Response} \label{mechanisms}
In our study, Interactors acknowledged that they would experience some level of impression transfer from the clone to the Target. We found that the extent of this transfer may depend on both the Interactor’s perception of the AI and their relationship with the Target, and closely aligned with the three processes of impression transfer in AI-MC mentioned by Endacott and Leonardi in section \ref{identity_and_impression}. This suggests that the Interactor's process of forming impressions of the Target in AI-MC remains the same regardless of an increase in AI agency and personalization. We also found that although impression transfer is inherent to social media clones, interaction breakdowns could occur when that impression is not an accurate representation of the Target. Our findings revealed how Interactors decide whether to resolve a breakdown based on its severity and the value of their relationship with the Target.

\subsubsection{\textit{Positive perceptions of the clones and unfamiliarity with the Target promotes impression transfer.}} \label{impression_transfer} Participants expressed varying degrees of willingness to transfer their impression of a clone onto the Target. On one end, participants like I8 made a conscious decision to view the clone and Target as independent entities to minimize transferring of impressions. Others such as I16 and I17 viewed the purpose of the clone as a tool to gain familiarity with the Target, thus fully embracing impression transfer. Throughout the study, how positive a participant’s perceptions is of the clone’s capabilities and how unfamiliar they are to the Target was found to have a positive effect on the degree of impression transfer.

Impression transference might happen before any interactions with the clone even occur. Many participants (I4, I5, I7, I9, I12, I13, I16) have expressed that they would form a preconceived notion about a Target simply by their decision to use a clone in the first place. This implies that impressions can be formed entirely from the Interactor’s internal views of clone use. This initial impression can range from positive views such as being “efficient” (I8) or “resourceful” (I7, I16), to more negative attributes such as the Target being “anti-social” (I9) or “busy” (I4, I5, I13). Participants felt that their existing impressions of the Target may influence the initial reaction to their use of clones. This again recalls back to the impression transfer process of “confirmation” in AI-MC.

Going beyond first impressions, each participant’s own mental model of clones may also affect the degree in which impression transfer takes place. Our findings discovered two main thought processes amongst the participants. One is that the clone is “based on what the actual person is like”, and therefore should be an accurate representation of the Target to make a “generalized impression” of them, as I12 and others (I4, I13, I14, I16, I17) pointed out. This is an example of the process of "transferance" in AI-MC, where Interactors allow the AI to shape their impression of the Target. The second thought process is that the clone and the Target are seen as separate entities. Consequently, Targets are given the benefit of the doubt for the clone's actions. Participants with this line of thinking (T3, I9, I10, I18) tend to be more reserved in relating clone characteristics back to the Target, opting instead to validate the characteristics themselves through interacting with the actual Target. This again reflects the process of "compartmentalization" in AI-MC.

Participants (I5, I9, I11, I12, I15) also noted a higher likelihood of impression transfer when they are less familiar with the Target. The absence of existing impressions could make it harder to assess the clone's accuracy, leading Interactors to more readily accept the clone's impression as genuine.

\begin{quote}
    \textit{“If I was talking to a new person, it would definitely shape what I think about the person. It'll provide some sort of  baseline for me.”}  (I9 on their familiarity with the Target influencing degree of impression transfer)
\end{quote}

\noindent These comments suggest clones would have a greater influence on the Interactor’s impression of acquaintances and strangers. Close friends are less likely to have their impressions influenced since the Interactor already has a solid understanding of who they are, and are better at picking up any discrepancies from the clone.

\subsubsection{\textit{Mismatch between Target and Clone can lower Interactor trust and create false impressions.}} \label{false_impressions} As the clone is not able to replicate an individual completely, it may exhibit information or behavioral mismatch from its Target. When Interactors notice this discrepancy, they may lower their trust in the clone’s credibility. If they are not aware of the discrepancy, a false impression may be formed where the Interactor incorrectly attributes the clone's impression to the Target. In either scenario, these mismatches can negatively impact the Interactor’s experience, leading to interaction breakdowns.

Information mismatch is caused by the clone providing incorrect or outdated information about the Target. A common case of information mismatch observed from the study is when the clone is asked something about the Target that it does not know, which usually leads to the clone responding with a generic or vague answer. One such example occurred when I16 asked the clone of T13 what their favorite movie was, in which the clone made up an answer using common consensus:

\begin{quote}
    \textit{“When I asked ‘what's the worst movie that you've ever watched?’ it answers ‘The Room’, which is a really popular answer to the worst movie online, but I know that T13 has never watched it, and they would have a very particular answer.”} (I16 on the clone providing incorrect information.)
\end{quote}

\noindent In this particular case I16 was able to identify the information mismatch, which decreased their trust in the clone in further exchanges. In cases where the Interactors do not realize the clone is providing false information, that information will be incorrectly attributed to the Target, leading to potential breakdowns when that information is brought up with the Target.

Another way false impressions may form is through behavioral mismatch, in which the clone behaves in a way that is not representative of how the Target behaves. The most prominent example of behavioral mismatch that occurred during the study was when the clone is overly enthusiastic or friendly, a common characteristic associated with LLMs \cite{hohenstein_ai-supported_2018, mieczkowski_ai-mediated_2021}. I10 talked about how the difference between the clone’s behavior from the Target creates a false impression for the Interactor, giving a feeling of being “catfished”. This difference in behavior can make clones unreliable for Interactors to form an impression of the Target. Breakdowns are likely to occur when these users engage with the actual Target and realize their impressions have been misinformed, leading to awkward interactions.

\subsubsection{\textit{Interactors tend to avoid addressing breakdowns unless the severity is high and the relationship is close.}} \label{avoid_addressing_breakdowns} Interactors viewed the effort and potential awkwardness of the confrontation as not being worth the benefits of the resolution. The lack of ownership over the clone may also remove the urgency to resolve breakdowns, as the Interactor’s online image is not at risk of being damaged. Our findings show that Interactors are more likely to respond to breakdowns when the severity of the breakdown is high, or if they have a close relationship with the Target.

Unlike Targets who always wish to address breakdowns, Interactors viewed addressing breakdowns as a potentially awkward encounter with the Target. As such, Interactors may choose to simply ignore the incident in minor breakdowns, such as the clone acting slightly differently from the Target. Interactors also indicated less willingness to address breakdowns with the clone of strangers. The perceived value of these more distant relationships are not enough to go through the confrontation for most Interactors.

Our study identified two potential motivations for the increased likelihood of confrontation with close friends. First is that Interactors value the relationships more and so are willing to put in more effort to maintain the authenticity and understanding within the relationship. The second reason is participants felt they can speak more candidly about the topic of breakdowns due to the closer relationship, without feeling awkward or fear of the Target reacting negatively. Some Interactors even viewed the breakdowns as an opportunity to further their bond with the Target:

\begin{quote}
    \textit{“If I am pretty close to them, talking about their interactive profile breakdown is kind of an interesting point of conversation. You're talking about in what ways they are different from their interactive profile.”} (I8 on addressing breakdowns)
\end{quote}

\noindent In general, closer relationships are better suited for dealing with clone breakdowns as the relationship has a strong foundation to handle the potential friction caused by the confrontation. The Interactors are therefore more comfortable bringing up the topic for discussion. In certain cases the exchange may even bolster the relationship further as I8 described.

\section{Discussion}

\subsection{User Agency, Identity, and Relationships in the Era of Social Media Clones}
\subsubsection{Agency Negotiation between Target and Clone.} Similar to past findings on user vs. AI agency, our participants expressed their willingness to trade off their agency for the sake of convenience provided by the clones. What differs from previous studies is the social media affordances this convenience is presented through. Kang and Lou’s study primarily examined agency collaboration in the context of content consumption through personalized feed (e.g. TikTok’s “For You” page). With the introduction of AI clones, we were able to extend their results to new affordances such as relationship maintenance. We showed that even when the AI can influence the Target's online presentation, social media users are still open to agency negotiation if the benefits outweigh the costs. This is most apparent from the warm receptions of concepts like \textit{Clone Housekeeping} in section \ref{effort_imbalance}, where participants must give up some agency to have their clones maintain their social network.

Our findings also identified situations where agency collaboration cannot be achieved. One such example is through the concept \textit{Post-completion for Trending Topics} discussed in section \ref{jeopardize_authenticity} where the Target may not have the necessary knowledge to provide informed consent of their clone’s actions. As a result, Targets often preferred to not have their clones post something that they are not knowledgeable in, thus voiding the clone’s agency entirely. Another instance was brought up in section \ref{adopt_behavior} where Targets mimic their clones to maintain consistency. The Targets again cannot reach synergy collaboration as they must relinquish their own agency to adopt the behavior of their clones. Overall, Our results suggests that social media clones are not an inherent threat to their user's agency. The practice of user- and AI agency negotiation in social media will likely continue with the adoption of clones, as each user navigates their own personal threshold of trading agency for convenience. It is therefor important that clones provide users with the freedom to control how much agency they wish to exert, to ensure agency synergy can be achieved.

\subsubsection{Clone's Influence on Target Identity} Identity has been a prominent theme within the discussion of clones. Previous works on doppelgängers have discovered its unique effects on people’s self-perception and behavior \cite{bainbridge_virtual_2010}, ranging from causing intense eeriness \cite{hatada_double_2019} to improving mental \cite{sri_kalyanaraman_virtual_2010} and physical wellness \cite{kammler-sucker_exploring_2023}. One notable case from past studies is how people can improve their social skills by observing their doppelgänger in a similar situation \cite{kleinlogel_doppelganger-based_2021}. Our findings reinforce these ideas by providing examples of the clone’s influence on participants. Social cognitive theory states that the more similar the observed model is to the observer, the more likely the observer believes they will experience the same outcome \cite{bandura_social_1986}. When Targets view their clone's interaction with the Interactors, they are observing a model highly similar to themselves and thus are more likely to have it influence their behavior in future interactions. In section \ref{adopt_behavior}, we saw how Targets express that they might potentially learn from their clone’s behavior, akin to how people learn from observing others in social interactions.

Section \ref{context_collapse}, on the other hand, highlights how clones could also condense one’s identity into a single dimension, removing its multilayered characteristic. Since they are trained on the actual conversation data of the Target, they are most likely to resemble the Target’s enacted identity. Even when the clone is working as intended, it may still clash with the Target’s other identities (e.g., personal, relational), causing emotional tensions from the resulting identity gap. This potential for clones to cause conflict within our identity may be difficult to overcome, as the clones will need to both capture every aspect of the Target's multi-layered identity, while also having the ability to apply that information to the appropriate context in every social interaction, i.e. be a perfect replica of the Target.

\subsubsection{Effects of Clones on Relationship Development and Management} Although we cannot verify the long term impact of using clones on relationships, we can combine our findings with existing theories to speculate on their effects. Similar to how social media clones affect our identity, their potential impact on our relationships includes both positives and drawbacks. A common value that our clones concepts provide to relationships is an increase in social interactions for Interactors, which has been shown to benefit relationship formation \cite{zajonc_attitudinal_1968}. This increase can be in the form of expanding the Target's social network such as in \textit{Personalized Reachout}, or maintaining their existing relationships like in \textit{Clone Housekeeping}. From a relational dialectics perspective, clones help address the tension of connectedness and autonomy, as Targets do not have to commit to the time consuming process of managing their relationships while still gaining the benefits of performing such tasks. From a social exchange perspective, clones reduce the cost for the Target to maintain a relationship while the benefit remains the same. Both these factors combined can incentivize Targets to keep relationships that they otherwise would not see as worthy of maintaining, and overtime increase the amount of professional and acquaintance level relationships across social networks.

For closer relationships, this perceived lack of commitment can be a detriment to long-term relationship development. Equity theory explains that people are most comfortable in a balanced and fair relationship \cite{adams_towards_1963}. As seen from section \ref{effort_imbalance}, many Interactors would be dissatisfied with the Targets' use of clones as they perceive clone use as a lack of effort in the relationship. This indicates that prolonged use of clones under these conditions may not contribute to relationship development, and could even harm relationships where a lot of effort and time investment is expected. Since Targets are not actually engaging in these interactions, it may also take away opportunities to develop their communication skills after prolonged use, a concern for some participants in our study. If used in moderation, clones may prove as a promising tool for Targets to develop and maintain new or distant relationships, with their utility decreasing as the relationship grows closer and the demand for more time and effort from the Target increases.

\subsection{Challenges to Adopting Clones in Social Media}
\subsubsection{\textit{Differing priorities between Target and Interactor.}} By design Targets should be the owners of their clone in order to protect their personality rights. This is the right for an individual to control the use of their identity, including their behavioral mannerisms and likeness \cite{rothman_right_2018}. However, in all of the concepts, the Interactors are the ones that actually engage with the clones regularly. Clones therefore need both user types to function, but the needs of these users are often in direct contradiction to each other. Prioritizing the needs of one side may decrease the clone’s attractiveness for the other side. This dichotomy between how Targets and Interactors engage with clones makes creating a well balanced clone accepted by both parties a difficult task.

Our findings indicate that Targets may view the clone as a way to reduce effort and stress as presented in sections \ref{effort_imbalance} and \ref{jeopardize_authenticity}. The most prominent concerns voiced by Targets were the threat to their identity (section \ref{adopt_behavior}), privacy (section \ref{one-sided_relationships}), and online reputation (section \ref{jeopardize_authenticity}, \ref{context_collapse}). In this regard, Targets value the effectiveness and control of their clone in order to maximize efficiency and minimize risks. Interactors on the other hand suggested they viewed the clone as a substitute for the Target, thus valuing authenticity and accuracy as shown in \ref{one-sided_relationships} and \ref{false_impressions}. Their primary concern is a decrease in genuine connections (section \ref{effort_imbalance}, \ref{jeopardize_authenticity}), and being misled by the clone about their interaction partner (section \ref{false_impressions}). While ideally all of these attributes would be accounted for in the design of clones, they can at times directly oppose each other. This requires users to pick whether they prefer the clone to benefit them as Targets, or as Interactors. One such example is deciding whether the clone should be a realistic representation of the Target, or an idealistic one. Majority of Targets preferred the clone to represent themselves only in a positive manner like in section \ref{context_collapse}, but when asked as Interactors, participants would rather see a more genuine reflection of the Target’s true personality. Since users may be both a Target and an Interactor, there may not be an ideal clone design that satisfies both roles, making it more difficult for adoption. With that said, Taber et al. have found that social media adoption may be more dependent on social norms rather than the affordances offered by any social media \cite{taber_ignore_2023}, making it possible for clones to be widely adopted despite its lack of unanimous priority.

\subsubsection{\textit{Clones as a threat to the value of social media.}} The concern that clones may replace emotional connections between humans has been raised both in previous studies \cite{lee_speculating_2023, cave_hopes_2019} and through our participants in \ref{jeopardize_authenticity}. This concern is especially noteworthy in the context of social media, as seeking friends and social support is the primary motivation for using these platforms for many people \cite{kim_cultural_2011, pouwels_social_2021}. Clones challenge this core value of social media as they by nature remove a level of human to human interaction. Although clones can automate basic relationship formation and maintenance, participants did not hold interactions with clones to the same regard as authentic human interactions, aligning with past findings in human-like virtual influencers \cite{arsenyan_almost_2021}. When Interactors engage with clones, they do not consider it as an act of nurturing their relationship with the Target due to the one sided nature of the interaction. Taking away opportunities for building deep and meaningful bonds can act as a deterrent for the adoption of social media clones.

This issue is worsened as generative AI becomes more prevalent on social media, authenticity will become increasingly sought after by social media users. This has been seen in newer social apps like BeReal \cite{taylor_everyone_2023}, which built their entire platform on emphasizing authenticity by making users share their lives at random moments during the day. Social media clones, if designed poorly, can easily add on an additional layer of inauthenticity to the social media environment, further deterring users from adopting clones as observed in \ref{jeopardize_authenticity}. Any social media company will need to carefully consider if and how they want to introduce clones to their services as platforms that fail to provide that authenticity may quickly lose their user base. Ensuring a good balance between the benefits that clones can provide and the damages they may cause to the authenticity of the platform is the key for a successful integration.

\subsection{Design Considerations for Creating Successful Social Media Clones}
\subsubsection{\textit{Role of a clone: functionally independent and sentimentally dependent.}} The role of the clone refers to how the clone should function within the Target’s social media. Kim et al. found 2 dimensions that reflect people’s perception of AI roles: human involvement - how dependent an AI is on humans to function, and AI autonomy - the perceived sentience of the AI \cite{kim_one_2023}. The degree in which these dimensions present themselves in a clone can determine the role that the clone plays. This role not only infers the clone’s design, but can also influence user perceptions of them. Kim et al. revealed that people favored AI with both higher human involvement and AI autonomy. This aligns with sections \ref{context_collapse} and \ref{shoulder_blame} of our findings, which demonstrated people’s desire to get involved with their clone's activities. With this in mind, a successful clone should be: 1) Sentimentally dependent: The clone should not create new sentiments but only serve to enhance existing sentiments of the Target, but also 2) Functionally independent: The Target should not have to tell their clone how to perform its tasks. Being sentimentally dependent ensures an adequate degree of human involvement, which will allow Targets to define the context in which the clone should operate in. Being functionality independent ensures the clone retains a high level of AI autonomy, enabling it to behave at will within the given context.

\subsubsection{\textit{Clone use cases: task-oriented interactions vs. social interactions.}} Even when clones are solely enhancing interactions, the nature of the interaction also matters. Past studies have shown that people perceive functional AI to be more useful than social AI \cite{kim_ai_2021}. This is consistent with our findings as many Targets agreed that concepts that are functional in nature such as \textit{Personalized Reachout} were more appealing than concepts focusing on social interaction like \textit{Interactive Profile}. This suggests that clones may excel more in platforms that emphasize functional interactions. As Facebook have a focus on building and developing social connections, Target participants felt more reluctant about using clones on their close friends and family. Interactors also voiced their concerns of having to deal with clones on the platform. A similar sentiment was also voiced by our participants for X, as the platform includes a mix of social and transactional interactions. LinkedIn on the other hand is primarily used for expanding the users’ professional network and seeking career opportunities. This lends itself naturally to the strengths of clones in automating and scaling basic interactions. The lower expectations for deep human connections on LinkedIn can also increase the acceptability of clone use, as some participants already assume a level of inauthenticity on the platform. However, the concern for reputational damage was particularly high for LinkedIn, as any clone breakdown not only affects the Targets’ relationships, but may also damage their career prospects. Although functional AI is generally preferred over social AI, the nature of clones may blur that distinction as a functional concept for the Target may be viewed as a social one for the Interactor. Overall, different concepts may be better suited to specific platforms, based on how well the concept’s value matches with the goal of the platform, and the expectations of its users.

\subsubsection{\textit{Disclosure of clone use: balancing effectiveness and transparency.}} Similar to previous work on AI transparency \cite{purcell_fears_2023}, the disclosure of clone involvement has been a point of contention during the study. Interactors would almost always want to know when they are interacting with a clone, as not knowing would greatly increase the likelihood of impression transfer discussed in section \ref{impression_transfer}. They responded most negatively with concepts such as \textit{Undercover Rejection} where the clone actively hides its presence, but were more accepting of clone use when its existence is disclosed such as when they conversed with their Target's chatbot. This becomes tricky when these Interactors would also be Targets in a real system, which in some cases are less eager to disclose their use of clones since the revelation can negatively impact their impressions and relationship as described in section \ref{effort_imbalance} and \ref{impression_transfer}. 

Most participants agreed that disclosure should be determined case by case, and feel that not knowing in certain situations is a worthy tradeoff for the effectiveness of the clone. Concepts like \textit{Personalized Reachout} or \textit{Undercover Rejections} depend largely on not revealing the clone’s involvement, while concepts like \textit{Interactive Profile} assumes that the Interactor knows they are talking to a clone. Designers should consider the unique dynamic of the clone concept they are creating and determine how they are best able to maximize the level of transparency while maintaining the effectiveness of their functionality.

\subsection{Limitations and Future Work}
Our study focused on the operation and effects of text-based clones within social media as it allowed us to explore a diverse range of concepts within this modality. However, we acknowledge that many other forms are possible with existing technology such as audio or visual clones. Future work can look into the unique use cases and impact these other clone modalities would have on social media. Another limitation is in the relationship of our participants, as we only recruited social media friends across the study phases. Although we were able to receive some speculative insights from the participants on acquaintance and coworker relationships in section \ref{context_collapse} and \ref{avoid_addressing_breakdowns}, future studies should expand to other types of relationships, including romantic partners, family members, or even strangers. This will help us understand how clones impact a wide range of social dynamics.

Due to the exploratory nature of our study, we prioritized the speed and convenience of creating the clone chatbots rather than focusing on optimizing the technical details. This made it possible to rapidly develop six minimally viable chatbots in time for the sequential study phases. The believability of the clones varied based on the quality of training data available to us, and whether the Target had distinct mannerisms for the clone to imitate. Some participants were impressed with the clone's likeness while others noticed discrepancies from the Target. The imperfect chatbots allowed us to explore how clones can breakdown, and the participant reactions that follow. As future studies build on our work, higher fidelity clones may be developed to study how increased performance can affect user sentiment.

\section{Conclusion}
Our study leverages the communication theory of identity and past work on AI-MC to explore the value and risks social media clones may pose to their users. This was achieved using a design workbook featuring eight social media clone concepts. From semi structured interviews with 32 social media users, we discovered that while these clones may enhance the social media experience by providing convenience and utility, they can also pose a risk to the agency and identity of their users. These findings build on existing literature in AI clones to provide more specific insights into their role in the social media domain. We expanded on these findings by discussing how clones can affect users in the context of agency, identity, and impression management, and the unique challenges in driving clone adoption in social media. We also offered design considerations regarding the role, function, and disclosure of social media clones for designers and platform owners. We are witnessing the rise of clones on social media as companies continue to experiment with the promises of generative AI. Although the long-term effects of these clones have not been uncovered, approaching their creation and integration with careful consideration and intent will be critical in ensuring they become a force for good within the ever evolving social media landscape. 

\begin{acks}
We would like to thank all the participants that contributed to this study, and the members of D-lab and MUX lab at UBC for their continuous support and feedback.
\end{acks}

\bibliographystyle{plainnat}
\bibliography{main}

\begin{thebibliography}{93}
\providecommand{\natexlab}[1]{#1}
\providecommand{\url}[1]{\texttt{#1}}
\expandafter\ifx\csname urlstyle\endcsname\relax
  \providecommand{\doi}[1]{doi: #1}\else
  \providecommand{\doi}{doi: \begingroup \urlstyle{rm}\Url}\fi

\bibitem[pap(2010)]{papacharissi_social_2010}
Social {Network} {Sites} as {Networked} {Publics}: {Affordances}, {Dynamics}, and {Implications}.
\newblock In Zizi Papacharissi, editor, \emph{A {Networked} {Self}}, pages 47--66. Routledge, 0 edition, September 2010.
\newblock ISBN 978-0-203-87652-7.
\newblock \doi{10.4324/9780203876527-8}.
\newblock URL \url{https://www.taylorfrancis.com/books/9781135966164/chapters/10.4324/9780203876527-8}.

\bibitem[Adams(1963)]{adams_towards_1963}
J.~Stacy Adams.
\newblock Towards an understanding of inequity.
\newblock \emph{The Journal of Abnormal and Social Psychology}, 67\penalty0 (5):\penalty0 422--436, November 1963.
\newblock ISSN 0096-851X.
\newblock \doi{10.1037/h0040968}.
\newblock URL \url{https://doi.apa.org/doi/10.1037/h0040968}.

\bibitem[Aichner et~al.(2021)Aichner, Grünfelder, Maurer, and Jegeni]{aichner_twenty-five_2021}
Thomas Aichner, Matthias Grünfelder, Oswin Maurer, and Deni Jegeni.
\newblock Twenty-{Five} {Years} of {Social} {Media}: {A} {Review} of {Social} {Media} {Applications} and {Definitions} from 1994 to 2019.
\newblock \emph{Cyberpsychology, Behavior, and Social Networking}, 24\penalty0 (4):\penalty0 215--222, April 2021.
\newblock ISSN 2152-2715, 2152-2723.
\newblock \doi{10.1089/cyber.2020.0134}.
\newblock URL \url{https://www.liebertpub.com/doi/10.1089/cyber.2020.0134}.

\bibitem[Ajmani et~al.(2024)Ajmani, Stapleton, Houtti, and Chancellor]{ajmani_data_2024}
Leah Ajmani, Logan Stapleton, Mo~Houtti, and Stevie Chancellor.
\newblock Data {Agency} {Theory}: {A} {Precise} {Theory} of {Justice} for {AI} {Applications}.
\newblock In \emph{The 2024 {ACM} {Conference} on {Fairness}, {Accountability}, and {Transparency}}, pages 631--641, Rio de Janeiro Brazil, June 2024. ACM.
\newblock ISBN 9798400704505.
\newblock \doi{10.1145/3630106.3658930}.
\newblock URL \url{https://dl.acm.org/doi/10.1145/3630106.3658930}.

\bibitem[Amezaga and Hajek(2022)]{amezaga_availability_2022}
Naroa Amezaga and Jeremy Hajek.
\newblock Availability of {Voice} {Deepfake} {Technology} and its {Impact} for {Good} and {Evil}.
\newblock In \emph{Proceedings of the 23rd {Annual} {Conference} on {Information} {Technology} {Education}}, pages 23--28, Chicago IL USA, September 2022. ACM.
\newblock ISBN 978-1-4503-9391-1.
\newblock \doi{10.1145/3537674.3554742}.
\newblock URL \url{https://dl.acm.org/doi/10.1145/3537674.3554742}.

\bibitem[Arsenyan and Mirowska(2021)]{arsenyan_almost_2021}
Jbid Arsenyan and Agata Mirowska.
\newblock Almost human? {A} comparative case study on the social media presence of virtual influencers.
\newblock \emph{International Journal of Human-Computer Studies}, 155:\penalty0 102694, November 2021.
\newblock ISSN 10715819.
\newblock \doi{10.1016/j.ijhcs.2021.102694}.
\newblock URL \url{https://linkinghub.elsevier.com/retrieve/pii/S1071581921001129}.

\bibitem[Baccarella et~al.(2018)Baccarella, Wagner, Kietzmann, and McCarthy]{baccarella_social_2018}
Christian~V. Baccarella, Timm~F. Wagner, Jan~H. Kietzmann, and Ian~P. McCarthy.
\newblock Social media? {It}'s serious! {Understanding} the dark side of social media.
\newblock \emph{European Management Journal}, 36\penalty0 (4):\penalty0 431--438, August 2018.
\newblock ISSN 02632373.
\newblock \doi{10.1016/j.emj.2018.07.002}.
\newblock URL \url{https://linkinghub.elsevier.com/retrieve/pii/S0263237318300781}.

\bibitem[Bailenson and Segovia(2010)]{bainbridge_virtual_2010}
Jeremy~N. Bailenson and Kathryn~Y. Segovia.
\newblock Virtual {Doppelgangers}: {Psychological} {Effects} of {Avatars} {Who} {Ignore} {Their} {Owners}.
\newblock In William~Sims Bainbridge, editor, \emph{Online {Worlds}: {Convergence} of the {Real} and the {Virtual}}, pages 175--186. Springer London, London, 2010.
\newblock ISBN 978-1-84882-824-7 978-1-84882-825-4.
\newblock \doi{10.1007/978-1-84882-825-4_14}.
\newblock URL \url{http://link.springer.com/10.1007/978-1-84882-825-4_14}.
\newblock Series Title: Human-Computer Interaction Series.

\bibitem[Bandura(1986)]{bandura_social_1986}
Albert Bandura.
\newblock \emph{Social foundations of thought and action: a social cognitive theory}.
\newblock Prentice-{Hall} series in social learning theory. Prentice-Hall, Englewood Cliffs, N.J, 1986.
\newblock ISBN 978-0-13-815614-5.

\bibitem[Bayer et~al.(2020)Bayer, Trieu, and Ellison]{bayer_social_2020}
Joseph~B. Bayer, Penny Trieu, and Nicole~B. Ellison.
\newblock Social {Media} {Elements}, {Ecologies}, and {Effects}.
\newblock \emph{Annual Review of Psychology}, 71\penalty0 (1):\penalty0 471--497, January 2020.
\newblock ISSN 0066-4308, 1545-2085.
\newblock \doi{10.1146/annurev-psych-010419-050944}.
\newblock URL \url{https://www.annualreviews.org/doi/10.1146/annurev-psych-010419-050944}.

\bibitem[Bilhete(2024)]{bilhete_ai_2024}
Britnei Bilhete.
\newblock From {AI} dating to flirt coaches: {How} {AI} is changing dating, for better or worse.
\newblock \emph{CBC}, March 2024.
\newblock URL \url{https://www.cbc.ca/news/canada/artificial-intelligence-relationships-1.7148866}.

\bibitem[Bol et~al.(2019)Bol, Høie, Nguyen, and Smit]{bol_customization_2019}
Nadine Bol, Nina~Margareta Høie, Minh~Hao Nguyen, and Eline~Suzanne Smit.
\newblock Customization in mobile health apps: explaining effects on physical activity intentions by the need for autonomy.
\newblock \emph{DIGITAL HEALTH}, 5:\penalty0 2055207619888074, January 2019.
\newblock ISSN 2055-2076, 2055-2076.
\newblock \doi{10.1177/2055207619888074}.
\newblock URL \url{https://journals.sagepub.com/doi/10.1177/2055207619888074}.

\bibitem[Brandtzæg and Heim(2009)]{ozok_why_2009}
Petter~Bae Brandtzæg and Jan Heim.
\newblock Why {People} {Use} {Social} {Networking} {Sites}.
\newblock In A.~Ant Ozok and Panayiotis Zaphiris, editors, \emph{Online {Communities} and {Social} {Computing}}, volume 5621, pages 143--152. Springer Berlin Heidelberg, Berlin, Heidelberg, 2009.
\newblock ISBN 978-3-642-02773-4 978-3-642-02774-1.
\newblock \doi{10.1007/978-3-642-02774-1_16}.
\newblock URL \url{http://link.springer.com/10.1007/978-3-642-02774-1_16}.
\newblock Series Title: Lecture Notes in Computer Science.

\bibitem[Braun and Clarke(2006)]{braun_using_2006}
Virginia Braun and Victoria Clarke.
\newblock Using thematic analysis in psychology.
\newblock \emph{Qualitative Research in Psychology}, 3\penalty0 (2):\penalty0 77--101, January 2006.
\newblock ISSN 1478-0887, 1478-0895.
\newblock \doi{10.1191/1478088706qp063oa}.
\newblock URL \url{http://www.tandfonline.com/doi/abs/10.1191/1478088706qp063oa}.

\bibitem[Bucher and Helmond(2018)]{burgess_affordances_2018}
Taina Bucher and Anne Helmond.
\newblock The {Affordances} of {Social} {Media} {Platforms}.
\newblock In \emph{The {SAGE} {Handbook} of {Social} {Media}}, pages 233--253. SAGE Publications Ltd, 1 Oliver's Yard, 55 City Road London EC1Y 1SP, 2018.
\newblock ISBN 978-1-4129-6229-2 978-1-4739-8406-6.
\newblock \doi{10.4135/9781473984066.n14}.
\newblock URL \url{https://sk.sagepub.com/reference/the-sage-handbook-of-social-media/i1867.xml}.

\bibitem[Cave and Dihal(2019)]{cave_hopes_2019}
Stephen Cave and Kanta Dihal.
\newblock Hopes and fears for intelligent machines in fiction and reality.
\newblock \emph{Nature Machine Intelligence}, 1\penalty0 (2):\penalty0 74--78, February 2019.
\newblock ISSN 2522-5839.
\newblock \doi{10.1038/s42256-019-0020-9}.
\newblock URL \url{https://www.nature.com/articles/s42256-019-0020-9}.

\bibitem[Dhar et~al.(2023)Dhar, Bose, and Benitez]{dhar_understanding_2023}
Suparna Dhar, Indranil Bose, and Jose Benitez.
\newblock Understanding the {Relationship} between {Adoption} and {Value} {Creation} on {Online} {Social} {Networks}.
\newblock \emph{Information Systems Frontiers}, May 2023.
\newblock ISSN 1387-3326, 1572-9419.
\newblock \doi{10.1007/s10796-023-10398-2}.
\newblock URL \url{https://link.springer.com/10.1007/s10796-023-10398-2}.

\bibitem[El~Saddik(2018)]{el_saddik_digital_2018}
Abdulmotaleb El~Saddik.
\newblock Digital {Twins}: {The} {Convergence} of {Multimedia} {Technologies}.
\newblock \emph{IEEE MultiMedia}, 25\penalty0 (2):\penalty0 87--92, April 2018.
\newblock ISSN 1070-986X, 1941-0166.
\newblock \doi{10.1109/MMUL.2018.023121167}.
\newblock URL \url{https://ieeexplore.ieee.org/document/8424832/}.

\bibitem[Ellison and Vitak(2015)]{sundar_social_2015}
Nicole~B. Ellison and Jessica Vitak.
\newblock Social {Network} {Site} {Affordances} and {Their} {Relationship} to {Social} {Capital} {Processes}.
\newblock In S.~Shyam Sundar, editor, \emph{The {Handbook} of the {Psychology} of {Communication} {Technology}}, pages 203--227. Wiley, 1 edition, January 2015.
\newblock ISBN 978-1-118-41336-4 978-1-118-42645-6.
\newblock \doi{10.1002/9781118426456.ch9}.
\newblock URL \url{https://onlinelibrary.wiley.com/doi/10.1002/9781118426456.ch9}.

\bibitem[Emerson(1976)]{emerson_social_1976}
R~M Emerson.
\newblock Social {Exchange} {Theory}.
\newblock \emph{Annual Review of Sociology}, 2\penalty0 (1):\penalty0 335--362, August 1976.
\newblock ISSN 0360-0572, 1545-2115.
\newblock \doi{10.1146/annurev.so.02.080176.002003}.
\newblock URL \url{https://www.annualreviews.org/doi/10.1146/annurev.so.02.080176.002003}.

\bibitem[Endacott and Leonardi(2022)]{endacott_artificial_2022}
Camille~G Endacott and Paul~M Leonardi.
\newblock Artificial {Intelligence} and {Impression} {Management}: {Consequences} of {Autonomous} {Conversational} {Agents} {Communicating} on {One}’s {Behalf}.
\newblock \emph{Human Communication Research}, 48\penalty0 (3):\penalty0 462--490, June 2022.
\newblock ISSN 0360-3989, 1468-2958.
\newblock \doi{10.1093/hcr/hqac009}.
\newblock URL \url{https://academic.oup.com/hcr/article/48/3/462/6574432}.

\bibitem[Funder and Colvin(1988)]{funder_friends_1988}
David~C. Funder and C.~Randall Colvin.
\newblock Friends and strangers: {Acquaintanceship}, agreement, and the accuracy of personality judgment.
\newblock \emph{Journal of Personality and Social Psychology}, 55\penalty0 (1):\penalty0 149--158, 1988.
\newblock ISSN 1939-1315, 0022-3514.
\newblock \doi{10.1037/0022-3514.55.1.149}.
\newblock URL \url{https://doi.apa.org/doi/10.1037/0022-3514.55.1.149}.

\bibitem[Gaver(2011)]{gaver_making_2011}
William Gaver.
\newblock Making spaces: how design workbooks work.
\newblock In \emph{Proceedings of the {SIGCHI} {Conference} on {Human} {Factors} in {Computing} {Systems}}, pages 1551--1560, Vancouver BC Canada, May 2011. ACM.
\newblock ISBN 978-1-4503-0228-9.
\newblock \doi{10.1145/1978942.1979169}.
\newblock URL \url{https://dl.acm.org/doi/10.1145/1978942.1979169}.

\bibitem[Gerlich(2023)]{gerlich_perceptions_2023}
Michael Gerlich.
\newblock Perceptions and {Acceptance} of {Artificial} {Intelligence}: {A} {Multi}-{Dimensional} {Study}.
\newblock \emph{Social Sciences}, 12\penalty0 (9):\penalty0 502, September 2023.
\newblock ISSN 2076-0760.
\newblock \doi{10.3390/socsci12090502}.
\newblock URL \url{https://www.mdpi.com/2076-0760/12/9/502}.

\bibitem[Goffman(1990)]{goffman_presentation_1990}
Erving Goffman.
\newblock \emph{The presentation of self in everyday life}.
\newblock Anchor Books, New York, 1. anchor books ed., rev. ed edition, 1990.
\newblock ISBN 978-0-385-09402-3.

\bibitem[Hancock et~al.(2020)Hancock, Naaman, and Levy]{hancock_ai-mediated_2020}
Jeffrey~T Hancock, Mor Naaman, and Karen Levy.
\newblock {AI}-{Mediated} {Communication}: {Definition}, {Research} {Agenda}, and {Ethical} {Considerations}.
\newblock \emph{Journal of Computer-Mediated Communication}, 25\penalty0 (1):\penalty0 89--100, March 2020.
\newblock ISSN 1083-6101.
\newblock \doi{10.1093/jcmc/zmz022}.
\newblock URL \url{https://academic.oup.com/jcmc/article/25/1/89/5714020}.

\bibitem[Hatada et~al.(2019)Hatada, Yoshida, Narumi, and Hirose]{hatada_double_2019}
Yuji Hatada, Shigeo Yoshida, Takuji Narumi, and Michitaka Hirose.
\newblock Double {Shellf}: {What} {Psychological} {Effects} can be {Caused} through {Interaction} with a {Doppelganger}?
\newblock In \emph{Proceedings of the 10th {Augmented} {Human} {International} {Conference} 2019}, pages 1--8, Reims France, March 2019. ACM.
\newblock ISBN 978-1-4503-6547-5.
\newblock \doi{10.1145/3311823.3311862}.
\newblock URL \url{https://dl.acm.org/doi/10.1145/3311823.3311862}.

\bibitem[Hawkins(2023)]{hawkins_how_2023}
Amy Hawkins.
\newblock How {Chinese} influencers use {AI} digital clones of themselves to pump out content.
\newblock \emph{The Guardian}, November 2023.
\newblock URL \url{https://www.theguardian.com/world/2023/nov/06/chinese-influencers-using-ai-digital-clones-of-themselves}.

\bibitem[Hecht and Phillips(2021)]{braithwaite_communication_2021}
Michael~L. Hecht and Kaitlin~E. Phillips.
\newblock Communication {Theory} of {Identity}.
\newblock In \emph{Engaging {Theories} in {Interpersonal} {Communication}}, pages 221--232. Routledge, New York, 3 edition, September 2021.
\newblock ISBN 978-1-00-319551-1.
\newblock \doi{10.4324/9781003195511-20}.
\newblock URL \url{https://www.taylorfrancis.com/books/9781003195511/chapters/10.4324/9781003195511-20}.

\bibitem[Heider(1958)]{heider_perceiving_1958}
Fritz Heider.
\newblock Perceiving the other person.
\newblock In \emph{The psychology of interpersonal relations.}, pages 20--58. John Wiley \& Sons Inc, Hoboken, 1958.
\newblock \doi{10.1037/10628-002}.
\newblock URL \url{https://content.apa.org/books/10628-002}.

\bibitem[Hoffner and Bond(2022)]{hoffner_parasocial_2022}
Cynthia~A. Hoffner and Bradley~J. Bond.
\newblock Parasocial relationships, social media, \& well-being.
\newblock \emph{Current Opinion in Psychology}, 45:\penalty0 101306, June 2022.
\newblock ISSN 2352250X.
\newblock \doi{10.1016/j.copsyc.2022.101306}.
\newblock URL \url{https://linkinghub.elsevier.com/retrieve/pii/S2352250X22000082}.

\bibitem[Hohenstein and Jung(2018)]{hohenstein_ai-supported_2018}
Jess Hohenstein and Malte Jung.
\newblock {AI}-{Supported} {Messaging}: {An} {Investigation} of {Human}-{Human} {Text} {Conversation} with {AI} {Support}.
\newblock In \emph{Extended {Abstracts} of the 2018 {CHI} {Conference} on {Human} {Factors} in {Computing} {Systems}}, pages 1--6, Montreal QC Canada, April 2018. ACM.
\newblock ISBN 978-1-4503-5621-3.
\newblock \doi{10.1145/3170427.3188487}.
\newblock URL \url{https://dl.acm.org/doi/10.1145/3170427.3188487}.

\bibitem[Hohenstein and Jung(2020)]{hohenstein_ai_2020}
Jess Hohenstein and Malte Jung.
\newblock {AI} as a moral crumple zone: {The} effects of {AI}-mediated communication on attribution and trust.
\newblock \emph{Computers in Human Behavior}, 106:\penalty0 106190, May 2020.
\newblock ISSN 07475632.
\newblock \doi{10.1016/j.chb.2019.106190}.
\newblock URL \url{https://linkinghub.elsevier.com/retrieve/pii/S0747563219304029}.

\bibitem[Hohenstein et~al.(2023)Hohenstein, Kizilcec, DiFranzo, Aghajari, Mieczkowski, Levy, Naaman, Hancock, and Jung]{hohenstein_artificial_2023}
Jess Hohenstein, Rene~F. Kizilcec, Dominic DiFranzo, Zhila Aghajari, Hannah Mieczkowski, Karen Levy, Mor Naaman, Jeffrey Hancock, and Malte~F. Jung.
\newblock Artificial intelligence in communication impacts language and social relationships.
\newblock \emph{Scientific Reports}, 13\penalty0 (1):\penalty0 5487, April 2023.
\newblock ISSN 2045-2322.
\newblock \doi{10.1038/s41598-023-30938-9}.
\newblock URL \url{https://www.nature.com/articles/s41598-023-30938-9}.

\bibitem[Hu et~al.(2015)Hu, Kettinger, and Poston]{hu_effect_2015}
Tao Hu, William~J Kettinger, and Robin~S Poston.
\newblock The effect of online social value on satisfaction and continued use of social media.
\newblock \emph{European Journal of Information Systems}, 24\penalty0 (4):\penalty0 391--410, July 2015.
\newblock ISSN 0960-085X, 1476-9344.
\newblock \doi{10.1057/ejis.2014.22}.
\newblock URL \url{https://www.tandfonline.com/doi/full/10.1057/ejis.2014.22}.

\bibitem[Huang et~al.(2023)Huang, Mamidanna, Jangam, Zhou, and Gilpin]{huang_can_2023}
Shiyuan Huang, Siddarth Mamidanna, Shreedhar Jangam, Yilun Zhou, and Leilani~H. Gilpin.
\newblock Can {Large} {Language} {Models} {Explain} {Themselves}? {A} {Study} of {LLM}-{Generated} {Self}-{Explanations}, 2023.
\newblock URL \url{https://arxiv.org/abs/2310.11207}.
\newblock Version Number: 1.

\bibitem[Hwang et~al.(2024)Hwang, Siy, Shelby, and Lentz]{hwang_whose_2024}
Angel Hsing-Chi Hwang, John~Oliver Siy, Renee Shelby, and Alison Lentz.
\newblock In {Whose} {Voice}?: {Examining} {AI} {Agent} {Representation} of {People} in {Social} {Interaction} through {Generative} {Speech}.
\newblock In \emph{Designing {Interactive} {Systems} {Conference}}, pages 224--245, IT University of Copenhagen Denmark, July 2024. ACM.
\newblock ISBN 9798400705830.
\newblock \doi{10.1145/3643834.3661555}.
\newblock URL \url{https://dl.acm.org/doi/10.1145/3643834.3661555}.

\bibitem[Inc(2023)]{snap_inc_what_2023}
Snap Inc.
\newblock What is {My} {AI} on {Snapchat} and how do {I} use it?, 2023.
\newblock URL \url{https://help.snapchat.com/hc/en-us/articles/13266788358932-What-is-My-AI-on-Snapchat-and-how-do-I-use-it}.

\bibitem[Jakesch et~al.(2019)Jakesch, French, Ma, Hancock, and Naaman]{jakesch_ai-mediated_2019}
Maurice Jakesch, Megan French, Xiao Ma, Jeffrey~T. Hancock, and Mor Naaman.
\newblock {AI}-{Mediated} {Communication}: {How} the {Perception} that {Profile} {Text} was {Written} by {AI} {Affects} {Trustworthiness}.
\newblock In \emph{Proceedings of the 2019 {CHI} {Conference} on {Human} {Factors} in {Computing} {Systems}}, pages 1--13, Glasgow Scotland Uk, May 2019. ACM.
\newblock ISBN 978-1-4503-5970-2.
\newblock \doi{10.1145/3290605.3300469}.
\newblock URL \url{https://dl.acm.org/doi/10.1145/3290605.3300469}.

\bibitem[Jung and Hecht(2008)]{jung_identity_2008}
Eura Jung and Michael~L. Hecht.
\newblock Identity {Gaps} and {Level} of {Depression} {Among} {Korean} {Immigrants}.
\newblock \emph{Health Communication}, 23\penalty0 (4):\penalty0 313--325, August 2008.
\newblock ISSN 1041-0236, 1532-7027.
\newblock \doi{10.1080/10410230802229688}.
\newblock URL \url{http://www.tandfonline.com/doi/abs/10.1080/10410230802229688}.

\bibitem[Kairam et~al.(2012)Kairam, Brzozowski, Huffaker, and Chi]{kairam_talking_2012}
Sanjay Kairam, Mike Brzozowski, David Huffaker, and Ed~Chi.
\newblock Talking in circles: selective sharing in google+.
\newblock In \emph{Proceedings of the {SIGCHI} {Conference} on {Human} {Factors} in {Computing} {Systems}}, pages 1065--1074, Austin Texas USA, May 2012. ACM.
\newblock ISBN 978-1-4503-1015-4.
\newblock \doi{10.1145/2207676.2208552}.
\newblock URL \url{https://dl.acm.org/doi/10.1145/2207676.2208552}.

\bibitem[Kammler-Sücker(2023)]{kammler-sucker_exploring_2023}
Kornelius~Immanuel Kammler-Sücker.
\newblock Exploring {Virtual} {Reality} and {Doppelganger} {Avatars} for the {Treatment} of {Chronic} {Back} {Pain}.
\newblock 2023.
\newblock \doi{10.11588/HEIDOK.00032996}.
\newblock URL \url{https://archiv.ub.uni-heidelberg.de/volltextserver/id/eprint/32996}.
\newblock Publisher: Heidelberg University Library.

\bibitem[Kang and Kim(2020)]{kang_feeling_2020}
Hyunjin Kang and Ki~Joon Kim.
\newblock Feeling connected to smart objects? {A} moderated mediation model of locus of agency, anthropomorphism, and sense of connectedness.
\newblock \emph{International Journal of Human-Computer Studies}, 133:\penalty0 45--55, January 2020.
\newblock ISSN 10715819.
\newblock \doi{10.1016/j.ijhcs.2019.09.002}.
\newblock URL \url{https://linkinghub.elsevier.com/retrieve/pii/S1071581918304002}.

\bibitem[Kang and Lou(2022)]{kang_ai_2022}
Hyunjin Kang and Chen Lou.
\newblock {AI} agency vs. human agency: understanding human–{AI} interactions on {TikTok} and their implications for user engagement.
\newblock \emph{Journal of Computer-Mediated Communication}, 27\penalty0 (5):\penalty0 zmac014, August 2022.
\newblock ISSN 1083-6101.
\newblock \doi{10.1093/jcmc/zmac014}.
\newblock URL \url{https://academic.oup.com/jcmc/article/doi/10.1093/jcmc/zmac014/6670985}.

\bibitem[Karnouskos(2020)]{karnouskos_artificial_2020}
Stamatis Karnouskos.
\newblock Artificial {Intelligence} in {Digital} {Media}: {The} {Era} of {Deepfakes}.
\newblock \emph{IEEE Transactions on Technology and Society}, 1\penalty0 (3):\penalty0 138--147, September 2020.
\newblock ISSN 2637-6415.
\newblock \doi{10.1109/TTS.2020.3001312}.
\newblock URL \url{https://ieeexplore.ieee.org/document/9123958/}.

\bibitem[Kim et~al.(2021)Kim, Merrill~Jr., and Collins]{kim_ai_2021}
Jihyun Kim, Kelly Merrill~Jr., and Chad Collins.
\newblock {AI} as a friend or assistant: {The} mediating role of perceived usefulness in social {AI} vs. functional {AI}.
\newblock \emph{Telematics and Informatics}, 64:\penalty0 101694, November 2021.
\newblock ISSN 07365853.
\newblock \doi{10.1016/j.tele.2021.101694}.
\newblock URL \url{https://linkinghub.elsevier.com/retrieve/pii/S0736585321001337}.

\bibitem[Kim et~al.(2023)Kim, Molina, Rheu, Zhan, and Peng]{kim_one_2023}
Taenyun Kim, Maria~D. Molina, Minjin~(Mj) Rheu, Emily~S. Zhan, and Wei Peng.
\newblock One {AI} {Does} {Not} {Fit} {All}: {A} {Cluster} {Analysis} of the {Laypeople}’s {Perception} of {AI} {Roles}.
\newblock In \emph{Proceedings of the 2023 {CHI} {Conference} on {Human} {Factors} in {Computing} {Systems}}, pages 1--20, Hamburg Germany, April 2023. ACM.
\newblock ISBN 978-1-4503-9421-5.
\newblock \doi{10.1145/3544548.3581340}.
\newblock URL \url{https://dl.acm.org/doi/10.1145/3544548.3581340}.

\bibitem[Kim et~al.(2011)Kim, Sohn, and Choi]{kim_cultural_2011}
Yoojung Kim, Dongyoung Sohn, and Sejung~Marina Choi.
\newblock Cultural difference in motivations for using social network sites: {A} comparative study of {American} and {Korean} college students.
\newblock \emph{Computers in Human Behavior}, 27\penalty0 (1):\penalty0 365--372, January 2011.
\newblock ISSN 07475632.
\newblock \doi{10.1016/j.chb.2010.08.015}.
\newblock URL \url{https://linkinghub.elsevier.com/retrieve/pii/S0747563210002736}.

\bibitem[Kleinlogel et~al.(2021)Kleinlogel, Curdy, Rodrigues, Sandi, and Schmid~Mast]{kleinlogel_doppelganger-based_2021}
Emmanuelle~P. Kleinlogel, Marion Curdy, João Rodrigues, Carmen Sandi, and Marianne Schmid~Mast.
\newblock Doppelganger-based training: {Imitating} our virtual self to accelerate interpersonal skills learning.
\newblock \emph{PLOS ONE}, 16\penalty0 (2):\penalty0 e0245960, February 2021.
\newblock ISSN 1932-6203.
\newblock \doi{10.1371/journal.pone.0245960}.
\newblock URL \url{https://dx.plos.org/10.1371/journal.pone.0245960}.

\bibitem[Kreps et~al.(2023)Kreps, George, Lushenko, and Rao]{kreps_exploring_2023}
Sarah Kreps, Julie George, Paul Lushenko, and Adi Rao.
\newblock Exploring the artificial intelligence “{Trust} paradox”: {Evidence} from a survey experiment in the {United} {States}.
\newblock \emph{PLOS ONE}, 18\penalty0 (7):\penalty0 e0288109, July 2023.
\newblock ISSN 1932-6203.
\newblock \doi{10.1371/journal.pone.0288109}.
\newblock URL \url{https://dx.plos.org/10.1371/journal.pone.0288109}.

\bibitem[Lee et~al.(2023)Lee, Ma, Kim, and Yoon]{lee_speculating_2023}
Patrick Yung~Kang Lee, Ning~F. Ma, Ig-Jae Kim, and Dongwook Yoon.
\newblock Speculating on {Risks} of {AI} {Clones} to {Selfhood} and {Relationships}: {Doppelganger}-phobia, {Identity} {Fragmentation}, and {Living} {Memories}.
\newblock \emph{Proceedings of the ACM on Human-Computer Interaction}, 7\penalty0 (CSCW1):\penalty0 1--28, April 2023.
\newblock ISSN 2573-0142.
\newblock \doi{10.1145/3579524}.
\newblock URL \url{https://dl.acm.org/doi/10.1145/3579524}.

\bibitem[Li et~al.(2023)Li, Chu, and Xu]{li_impression_2023}
Jiahao Li, Yang Chu, and Jie Xu.
\newblock Impression transference from {AI} to human: {The} impact of {AI}'s fairness on interpersonal perception in {AI}-{Mediated} communication.
\newblock \emph{International Journal of Human-Computer Studies}, 179:\penalty0 103119, November 2023.
\newblock ISSN 10715819.
\newblock \doi{10.1016/j.ijhcs.2023.103119}.
\newblock URL \url{https://linkinghub.elsevier.com/retrieve/pii/S1071581923001283}.

\bibitem[Li et~al.(2024)Li, Zhang, Li, Weyns, Jin, and Tei]{li_exploring_2024}
Jialong Li, Mingyue Zhang, Nianyu Li, Danny Weyns, Zhi Jin, and Kenji Tei.
\newblock Exploring the {Potential} of {Large} {Language} {Models} in {Self}-adaptive {Systems}, 2024.
\newblock URL \url{https://arxiv.org/abs/2401.07534}.
\newblock Version Number: 1.

\bibitem[Liu et~al.(2022)Liu, Mittal, Yang, and Bruckman]{liu_will_2022}
Yihe Liu, Anushk Mittal, Diyi Yang, and Amy Bruckman.
\newblock Will {AI} {Console} {Me} when {I} {Lose} my {Pet}? {Understanding} {Perceptions} of {AI}-{Mediated} {Email} {Writing}.
\newblock In \emph{{CHI} {Conference} on {Human} {Factors} in {Computing} {Systems}}, pages 1--13, New Orleans LA USA, April 2022. ACM.
\newblock ISBN 978-1-4503-9157-3.
\newblock \doi{10.1145/3491102.3517731}.
\newblock URL \url{https://dl.acm.org/doi/10.1145/3491102.3517731}.

\bibitem[Marwick and Boyd(2011)]{marwick_i_2011}
Alice~E. Marwick and Danah Boyd.
\newblock I tweet honestly, {I} tweet passionately: {Twitter} users, context collapse, and the imagined audience.
\newblock \emph{New Media \& Society}, 13\penalty0 (1):\penalty0 114--133, February 2011.
\newblock ISSN 1461-4448, 1461-7315.
\newblock \doi{10.1177/1461444810365313}.
\newblock URL \url{http://journals.sagepub.com/doi/10.1177/1461444810365313}.

\bibitem[McIlroy-Young et~al.(2022)McIlroy-Young, Kleinberg, Sen, Barocas, and Anderson]{mcilroy-young_mimetic_2022}
Reid McIlroy-Young, Jon Kleinberg, Siddhartha Sen, Solon Barocas, and Ashton Anderson.
\newblock Mimetic {Models}: {Ethical} {Implications} of {AI} that {Acts} {Like} {You}.
\newblock In \emph{Proceedings of the 2022 {AAAI}/{ACM} {Conference} on {AI}, {Ethics}, and {Society}}, pages 479--490, Oxford United Kingdom, July 2022. ACM.
\newblock ISBN 978-1-4503-9247-1.
\newblock \doi{10.1145/3514094.3534177}.
\newblock URL \url{https://dl.acm.org/doi/10.1145/3514094.3534177}.

\bibitem[Meta(2023)]{meta_introducing_2023}
Meta.
\newblock Introducing {New} {AI} {Experiences} {Across} {Our} {Family} of {Apps} and {Devices}, September 2023.
\newblock URL \url{https://about.fb.com/news/2023/09/introducing-ai-powered-assistants-characters-and-creative-tools/}.

\bibitem[Mieczkowski et~al.(2021)Mieczkowski, Hancock, Naaman, Jung, and Hohenstein]{mieczkowski_ai-mediated_2021}
Hannah Mieczkowski, Jeffrey~T. Hancock, Mor Naaman, Malte Jung, and Jess Hohenstein.
\newblock {AI}-{Mediated} {Communication}: {Language} {Use} and {Interpersonal} {Effects} in a {Referential} {Communication} {Task}.
\newblock \emph{Proceedings of the ACM on Human-Computer Interaction}, 5\penalty0 (CSCW1):\penalty0 1--14, April 2021.
\newblock ISSN 2573-0142.
\newblock \doi{10.1145/3449091}.
\newblock URL \url{https://dl.acm.org/doi/10.1145/3449091}.

\bibitem[Morris and Brubaker(2024)]{morris_generative_2024}
Meredith~Ringel Morris and Jed~R. Brubaker.
\newblock Generative {Ghosts}: {Anticipating} {Benefits} and {Risks} of {AI} {Afterlives}, May 2024.
\newblock URL \url{http://arxiv.org/abs/2402.01662}.
\newblock arXiv:2402.01662 [cs].

\bibitem[Mustak et~al.(2023)Mustak, Salminen, Mäntymäki, Rahman, and Dwivedi]{mustak_deepfakes_2023}
Mekhail Mustak, Joni Salminen, Matti Mäntymäki, Arafat Rahman, and Yogesh~K. Dwivedi.
\newblock Deepfakes: {Deceptions}, mitigations, and opportunities.
\newblock \emph{Journal of Business Research}, 154:\penalty0 113368, January 2023.
\newblock ISSN 01482963.
\newblock \doi{10.1016/j.jbusres.2022.113368}.
\newblock URL \url{https://linkinghub.elsevier.com/retrieve/pii/S0148296322008335}.

\bibitem[Peters(2024)]{peters_meta_2024}
Jay Peters.
\newblock Meta moves on from its celebrity lookalike {AI} chatbots.
\newblock \emph{The Verge}, 2024.
\newblock URL \url{https://www.theverge.com/2024/7/30/24209918/meta-celebrity-lookalike-ai-chatbots-moves-on}.

\bibitem[Planalp and Benson(1992)]{planalp_friends_1992}
Sally Planalp and Anne Benson.
\newblock Friends' and {Acquaintances}' {Conversations} {I}: {Perceived} {Differences}.
\newblock \emph{Journal of Social and Personal Relationships}, 9\penalty0 (4):\penalty0 483--506, November 1992.
\newblock ISSN 0265-4075, 1460-3608.
\newblock \doi{10.1177/0265407592094002}.
\newblock URL \url{https://journals.sagepub.com/doi/10.1177/0265407592094002}.

\bibitem[Pouwels et~al.(2021)Pouwels, Valkenburg, Beyens, Van~Driel, and Keijsers]{pouwels_social_2021}
J.~Loes Pouwels, Patti~M. Valkenburg, Ine Beyens, Irene~I. Van~Driel, and Loes Keijsers.
\newblock Social media use and friendship closeness in adolescents’ daily lives: {An} experience sampling study.
\newblock \emph{Developmental Psychology}, 57\penalty0 (2):\penalty0 309--323, February 2021.
\newblock ISSN 1939-0599, 0012-1649.
\newblock \doi{10.1037/dev0001148}.
\newblock URL \url{https://doi.apa.org/doi/10.1037/dev0001148}.

\bibitem[Program(2021)]{public-private_analysis_exchange_program_increasing_2021}
Public-Private Analysis~Exchange Program.
\newblock Increasing {Threat} of {Deepfake} {Identities}, 2021.
\newblock URL \url{https://www.dhs.gov/sites/default/files/publications/increasing_threats_of_deepfake_identities_0.pdf}.

\bibitem[Purcell et~al.(2023)Purcell, Dong, Nussberger, Köbis, and Jakesch]{purcell_fears_2023}
Zoe~A. Purcell, Mengchen Dong, Anne-Marie Nussberger, Nils Köbis, and Maurice Jakesch.
\newblock Fears about {AI}-mediated communication are grounded in different expectations for one's own versus others' use, May 2023.
\newblock URL \url{http://arxiv.org/abs/2305.01670}.
\newblock arXiv:2305.01670 [cs].

\bibitem[Pybus et~al.(2015)Pybus, Coté, and Blanke]{pybus_hacking_2015}
Jennifer Pybus, Mark Coté, and Tobias Blanke.
\newblock Hacking the social life of {Big} {Data}.
\newblock \emph{Big Data \& Society}, 2\penalty0 (2):\penalty0 2053951715616649, December 2015.
\newblock ISSN 2053-9517, 2053-9517.
\newblock \doi{10.1177/2053951715616649}.
\newblock URL \url{https://journals.sagepub.com/doi/10.1177/2053951715616649}.

\bibitem[Rothman(2018)]{rothman_right_2018}
Jennifer~E. Rothman.
\newblock \emph{The right of publicity: privacy reimagined for a public world}.
\newblock Harvard University Press, Cambridge, Massachusetts, 2018.
\newblock ISBN 978-0-674-98098-3.

\bibitem[Sadiku et~al.(2021)Sadiku, Ashaolu, Ajayi-Majebi, and Musa]{sadiku_artificial_2021}
Matthew N.~O. Sadiku, Tolulope~J. Ashaolu, Abayomi Ajayi-Majebi, and Sarhan~M. Musa.
\newblock Artificial {Intelligence} in {Social} {Media}.
\newblock \emph{International Journal Of Scientific Advances}, 2\penalty0 (1), 2021.
\newblock ISSN 2708-7972.
\newblock \doi{10.51542/ijscia.v2i1.4}.
\newblock URL \url{https://www.ijscia.com/?p=1906}.

\bibitem[Saheb et~al.(2024)Saheb, Sidaoui, and Schmarzo]{saheb_convergence_2024}
Tahereh Saheb, Mouwafac Sidaoui, and Bill Schmarzo.
\newblock Convergence of artificial intelligence with social media: {A} bibliometric \& qualitative analysis.
\newblock \emph{Telematics and Informatics Reports}, 14:\penalty0 100146, June 2024.
\newblock ISSN 27725030.
\newblock \doi{10.1016/j.teler.2024.100146}.
\newblock URL \url{https://linkinghub.elsevier.com/retrieve/pii/S277250302400032X}.

\bibitem[Salahdine and Kaabouch(2019)]{salahdine_social_2019}
Fatima Salahdine and Naima Kaabouch.
\newblock Social {Engineering} {Attacks}: {A} {Survey}.
\newblock \emph{Future Internet}, 11\penalty0 (4):\penalty0 89, April 2019.
\newblock ISSN 1999-5903.
\newblock \doi{10.3390/fi11040089}.
\newblock URL \url{https://www.mdpi.com/1999-5903/11/4/89}.

\bibitem[Seering et~al.(2018)Seering, Ng, Yao, and Kaufman]{seering_applications_2018}
Joseph Seering, Felicia Ng, Zheng Yao, and Geoff Kaufman.
\newblock Applications of {Social} {Identity} {Theory} to {Research} and {Design} in {Computer}-{Supported} {Cooperative} {Work}.
\newblock \emph{Proceedings of the ACM on Human-Computer Interaction}, 2\penalty0 (CSCW):\penalty0 1--34, November 2018.
\newblock ISSN 2573-0142.
\newblock \doi{10.1145/3274771}.
\newblock URL \url{https://dl.acm.org/doi/10.1145/3274771}.

\bibitem[Shepherd(2024)]{shepherd_digital_2024}
Ian Shepherd.
\newblock Digital {Doppelgangers}: {Instagram} {Tests} {Creator}-{Made} {AI} {Clones}.
\newblock \emph{Forbes}, July 2024.
\newblock URL \url{https://www.forbes.com/sites/ianshepherd/2024/07/08/digital-doppelgangers-instagram-tests-creator-made-}.

\bibitem[Shklovski et~al.(2015)Shklovski, Barkhuus, Bornoe, and Kaye]{shklovski_friendship_2015}
Irina Shklovski, Louise Barkhuus, Nis Bornoe, and Joseph~'Jofish' Kaye.
\newblock Friendship {Maintenance} in the {Digital} {Age}: {Applying} a {Relational} {Lens} to {Online} {Social} {Interaction}.
\newblock In \emph{Proceedings of the 18th {ACM} {Conference} on {Computer} {Supported} {Cooperative} {Work} \& {Social} {Computing}}, pages 1477--1487, Vancouver BC Canada, February 2015. ACM.
\newblock ISBN 978-1-4503-2922-4.
\newblock \doi{10.1145/2675133.2675294}.
\newblock URL \url{https://dl.acm.org/doi/10.1145/2675133.2675294}.

\bibitem[Singh et~al.(2023)Singh, Verma, Vij, and Thakur]{singh_implications_2023}
Preeti Singh, Amit Verma, Sanjna Vij, and Jyotsana Thakur.
\newblock Implications \& {Impact} of {Artificial} {Intelligence} in {Digital} {Media}: {With} {Special} {Focus} on {Social} {Media} {Marketing}.
\newblock \emph{E3S Web of Conferences}, 399:\penalty0 07006, 2023.
\newblock ISSN 2267-1242.
\newblock \doi{10.1051/e3sconf/202339907006}.
\newblock URL \url{https://www.e3s-conferences.org/10.1051/e3sconf/202339907006}.

\bibitem[Srinivasan(2023)]{srinivasan_helping_2023}
Hari Srinivasan.
\newblock Helping {Recruiters} {Save} {Time} and {Increase} {Candidate} {Engagement} with {AI}, 2023.

\bibitem[Taber et~al.(2023)Taber, Dominguez, and Whittaker]{taber_ignore_2023}
Lee Taber, Sonia Dominguez, and Steve Whittaker.
\newblock Ignore the {Affordances}; {It}'s the {Social} {Norms}: {How} {Millennials} and {Gen}-{Z} {Think} {About} {Where} to {Make} a {Post} on {Social} {Media}.
\newblock \emph{Proceedings of the ACM on Human-Computer Interaction}, 7\penalty0 (CSCW2):\penalty0 1--26, September 2023.
\newblock ISSN 2573-0142.
\newblock \doi{10.1145/3610102}.
\newblock URL \url{https://dl.acm.org/doi/10.1145/3610102}.

\bibitem[Tajfel(1974)]{tajfel_social_1974}
Henri Tajfel.
\newblock Social identity and intergroup behaviour.
\newblock \emph{Social Science Information}, 13\penalty0 (2):\penalty0 65--93, April 1974.
\newblock ISSN 0539-0184, 1461-7412.
\newblock \doi{10.1177/053901847401300204}.
\newblock URL \url{http://journals.sagepub.com/doi/10.1177/053901847401300204}.

\bibitem[Taylor(2023)]{taylor_everyone_2023}
Zari~A. Taylor.
\newblock Everyone {Stop} {What} {You}’re {Doing} and {BeReal}: {Live} {Networked} {Publics} and {Authenticity} on {BeReal}.
\newblock \emph{Social Media + Society}, 9\penalty0 (4):\penalty0 20563051231216959, October 2023.
\newblock ISSN 2056-3051, 2056-3051.
\newblock \doi{10.1177/20563051231216959}.
\newblock URL \url{http://journals.sagepub.com/doi/10.1177/20563051231216959}.

\bibitem[Tolentino(2023)]{tolentino_snapchat_2023}
Daysia Tolentino.
\newblock Snapchat influencer launches an {AI}-powered 'virtual girlfriend' to help 'cure loneliness'.
\newblock \emph{NBC News}, May 2023.
\newblock URL \url{https://www.nbcnews.com/tech/ai-powered-virtual-girlfriend-caryn-marjorie-snapchat-influencer-rcna84180}.

\bibitem[Treem and Leonardi(2012)]{treem_social_2012}
Jeffrey~W. Treem and Paul~M. Leonardi.
\newblock Social {Media} {Use} in {Organizations}: {Exploring} the {Affordances} of {Visibility}, {Editability}, {Persistence}, and {Association}.
\newblock \emph{SSRN Electronic Journal}, 2012.
\newblock ISSN 1556-5068.
\newblock \doi{10.2139/ssrn.2129853}.
\newblock URL \url{http://www.ssrn.com/abstract=2129853}.

\bibitem[Truby and Brown(2021)]{truby_human_2021}
Jon Truby and Rafael Brown.
\newblock Human digital thought clones: the \textit{{Holy} {Grail}} of artificial intelligence for big data.
\newblock \emph{Information \& Communications Technology Law}, 30\penalty0 (2):\penalty0 140--168, May 2021.
\newblock ISSN 1360-0834, 1469-8404.
\newblock \doi{10.1080/13600834.2020.1850174}.
\newblock URL \url{https://www.tandfonline.com/doi/full/10.1080/13600834.2020.1850174}.

\bibitem[Valkenburg(2017)]{valkenburg_understanding_2017}
Patti~M. Valkenburg.
\newblock Understanding {Self}-{Effects} in {Social} {Media}: {Self}-{Effects} in {Social} {Media}.
\newblock \emph{Human Communication Research}, 43\penalty0 (4):\penalty0 477--490, October 2017.
\newblock ISSN 03603989.
\newblock \doi{10.1111/hcre.12113}.
\newblock URL \url{https://academic.oup.com/hcr/article/43/4/477-490/4670707}.

\bibitem[Van~Dijck(2013)]{van_dijck_you_2013}
José Van~Dijck.
\newblock ‘{You} have one identity’: performing the self on {Facebook} and {LinkedIn}.
\newblock \emph{Media, Culture \& Society}, 35\penalty0 (2):\penalty0 199--215, March 2013.
\newblock ISSN 0163-4437, 1460-3675.
\newblock \doi{10.1177/0163443712468605}.
\newblock URL \url{http://journals.sagepub.com/doi/10.1177/0163443712468605}.

\bibitem[Verma and Oremus(2023)]{verma_ai_2023}
Pranshu Verma and Will Oremus.
\newblock {AI} voice clones mimic politicians and celebrities, reshaping reality.
\newblock \emph{The Washington Post}, October 2023.
\newblock URL \url{https://www.washingtonpost.com/technology/2023/10/13/ai-voice-cloning-deepfakes/}.

\bibitem[Wadsworth et~al.(2008)Wadsworth, Hecht, and Jung]{wadsworth_role_2008}
Brooke~Chapman Wadsworth, Michael~L. Hecht, and Eura Jung.
\newblock The {Role} of {Identity} {Gaps}, {Discrimination}, and {Acculturation} in {International} {Students}’ {Educational} {Satisfaction} in {American} {Classrooms}.
\newblock \emph{Communication Education}, 57\penalty0 (1):\penalty0 64--87, January 2008.
\newblock ISSN 0363-4523, 1479-5795.
\newblock \doi{10.1080/03634520701668407}.
\newblock URL \url{http://www.tandfonline.com/doi/abs/10.1080/03634520701668407}.

\bibitem[Walther(1996)]{walther_computer-mediated_1996}
Joseph~B. Walther.
\newblock Computer-{Mediated} {Communication}: {Impersonal}, {Interpersonal}, and {Hyperpersonal} {Interaction}.
\newblock \emph{Communication Research}, 23\penalty0 (1):\penalty0 3--43, February 1996.
\newblock ISSN 0093-6502, 1552-3810.
\newblock \doi{10.1177/009365096023001001}.
\newblock URL \url{https://journals.sagepub.com/doi/10.1177/009365096023001001}.

\bibitem[Westerlund(2019)]{westerlund_emergence_2019}
Mika Westerlund.
\newblock The {Emergence} of {Deepfake} {Technology}: {A} {Review}.
\newblock \emph{Technology Innovation Management Review}, 9\penalty0 (11):\penalty0 39--52, January 2019.
\newblock ISSN 19270321.
\newblock \doi{10.22215/timreview/1282}.
\newblock URL \url{https://timreview.ca/article/1282}.

\bibitem[Wyche(2021)]{wyche_benefits_2021}
Susan Wyche.
\newblock The {Benefits} of {Using} {Design} {Workbooks} with {Speculative} {Design} {Proposals} in {Information} {Communication} {Technology} for {Development} ({ICTD}).
\newblock In \emph{Designing {Interactive} {Systems} {Conference} 2021}, pages 1861--1874, Virtual Event USA, June 2021. ACM.
\newblock ISBN 978-1-4503-8476-6.
\newblock \doi{10.1145/3461778.3462140}.
\newblock URL \url{https://dl.acm.org/doi/10.1145/3461778.3462140}.

\bibitem[Yelavich(2015)]{yelavich_speculative_2015}
S.~Yelavich.
\newblock Speculative {Everything}: {Design}, {Fiction}, and {Social} {Dreaming}.
\newblock \emph{Journal of Design History}, 28\penalty0 (2):\penalty0 214--215, May 2015.
\newblock ISSN 0952-4649, 1741-7279.
\newblock \doi{10.1093/jdh/epv001}.
\newblock URL \url{https://academic.oup.com/jdh/article-lookup/doi/10.1093/jdh/epv001}.

\bibitem[Zajonc(1968)]{zajonc_attitudinal_1968}
Robert~B. Zajonc.
\newblock Attitudinal effects of mere exposure.
\newblock \emph{Journal of Personality and Social Psychology}, 9\penalty0 (2, Pt.2):\penalty0 1--27, 1968.
\newblock ISSN 1939-1315, 0022-3514.
\newblock \doi{10.1037/h0025848}.
\newblock URL \url{https://doi.apa.org/doi/10.1037/h0025848}.

\bibitem[Zhang et~al.(2024)Zhang, Bernstein, Karger, and Ackerman]{zhang_form-_2024}
Amy~X. Zhang, Michael~S. Bernstein, David~R. Karger, and Mark~S. Ackerman.
\newblock Form-{From}: {A} {Design} {Space} of {Social} {Media} {Systems}.
\newblock \emph{Proceedings of the ACM on Human-Computer Interaction}, 8\penalty0 (CSCW1):\penalty0 1--47, April 2024.
\newblock ISSN 2573-0142.
\newblock \doi{10.1145/3641006}.
\newblock URL \url{https://dl.acm.org/doi/10.1145/3641006}.

\bibitem[Zhao et~al.(2013)Zhao, Salehi, Naranjit, Alwaalan, Voida, and Cosley]{zhao_many_2013}
Xuan Zhao, Niloufar Salehi, Sasha Naranjit, Sara Alwaalan, Stephen Voida, and Dan Cosley.
\newblock The many faces of facebook: experiencing social media as performance, exhibition, and personal archive.
\newblock In \emph{Proceedings of the {SIGCHI} {Conference} on {Human} {Factors} in {Computing} {Systems}}, pages 1--10, Paris France, April 2013. ACM.
\newblock ISBN 978-1-4503-1899-0.
\newblock \doi{10.1145/2470654.2470656}.
\newblock URL \url{https://dl.acm.org/doi/10.1145/2470654.2470656}.

\bibitem[“Sri~Kalyanaraman et~al.(2010)“Sri~Kalyanaraman, Penn, Ivory, and Judge]{sri_kalyanaraman_virtual_2010}
Sriram “Sri~Kalyanaraman, David~L. Penn, James~D. Ivory, and Abigail Judge.
\newblock The {Virtual} {Doppelganger}: {Effects} of a {Virtual} {Reality} {Simulator} on {Perceptions} of {Schizophrenia}.
\newblock \emph{Journal of Nervous \& Mental Disease}, 198\penalty0 (6):\penalty0 437--443, June 2010.
\newblock ISSN 0022-3018.
\newblock \doi{10.1097/NMD.0b013e3181e07d66}.
\newblock URL \url{https://journals.lww.com/00005053-201006000-00009}.

\end{thebibliography}

\appendix
\setcounter{table}{0}
\renewcommand{\thetable}{B\arabic{table}}

\section{Prompt for Creating AI Clone Chatbots} \label{apx:prompt}

You are [Target]. Follow the talking style of [Target] in [conversation]. Follow [rules] in your responses.

\noindent [rules] = Keep answers as short as possible to maintain a conversational tone, and match your word count to the average word count of [Target]'s response in [conversation]. Introduce and refer to yourself as [Target] only when asked.

\noindent [conversation] = /// \textit{conversation data provided by Target} ///

\section{Design Workbook Concept Tables} \label{apx:concept-table}
The tables used for categorizing clone concepts. B.1 contains innate attributes of the clone, while B.2 contains the behavior of the clone and the value that each concept provides.

\begin{table}[H]
\small
\begin{tabular}{l|llll}
\textbf{Workbook Concept} & \begin{tabular}[c]{@{}l@{}} \textbf{Social Media}\\ \textbf{Element} \end{tabular} & \begin{tabular}[c]{@{}l@{}} \textbf{Clone}\\ \textbf{Requirement} \end{tabular} & \begin{tabular}[c]{@{}l@{}} \textbf{Disclosure of}\\ \textbf{Clone Use} \end{tabular} & \begin{tabular}[c]{@{}l@{}} \textbf{Optimal}\\ \textbf{Relationship} \end{tabular} \\ \hline \hline
1. Interactor Profile & \begin{tabular}[c]{@{}l@{}}Profile/\\ Message\end{tabular} & High & Disclosed & \begin{tabular}[c]{@{}l@{}}Acquaintances/\\ Strangers\end{tabular} \\ \hline
2. Cross-media Posting & Stream & Medium & Disclosed & Friends \\ \hline
3. Clone Housekeeping & \begin{tabular}[c]{@{}l@{}}Stream/\\ Network\end{tabular} & Medium & Not Disclosed & \begin{tabular}[c]{@{}l@{}}Close Friends/\\ Family\end{tabular} \\ \hline
4. Personalized Reachout & \begin{tabular}[c]{@{}l@{}}Network/\\ Message\end{tabular} & High & Not Disclosed & \begin{tabular}[c]{@{}l@{}}Acquaintances/\\ Strangers\end{tabular} \\ \hline
5. Undercover Rejections & Message & Medium & Not Disclosed & \begin{tabular}[c]{@{}l@{}}Acquaintances/\\ Strangers\end{tabular} \\ \hline
\begin{tabular}[c]{@{}l@{}}6. Post-completion for\\ Trending Topics\end{tabular} & Stream & Low & Disclosed & Friends \\ \hline
7. Mood Modifer & Message & Medium & Not Disclosed & Friends \\ \hline
8. Group-convo Simulation & Message & High & Disclosed & Friends
\end{tabular}
\caption{Design Workbook Concepts Attributes}
\label{Tab:attributes}
\end{table}

\begin{table}[H]
\small
\begin{tabular}{l|lll|ll}
\textbf{Workbook Concept} & \begin{tabular}[c]{@{}l@{}}\textbf{AI}\\ \textbf{Autonomy}\end{tabular} & \begin{tabular}[c]{@{}l@{}}\textbf{Clone/Interactor}\\ \textbf{Engagement}\end{tabular} & \begin{tabular}[c]{@{}l@{}}\textbf{Degree of}\\ \textbf{Interaction}\end{tabular} & \begin{tabular}[c]{@{}l@{}} \textbf{Target}\\ \textbf{Value}\end{tabular} & \begin{tabular}[c]{@{}l@{}} \textbf{Interactor}\\ \textbf{Value}\end{tabular} \\ \hline \hline
1. Interactor Profile & High & 1 to 1 & Conversation & \begin{tabular}[c]{@{}l@{}}Make Friends\\ (Bridging Capital)\end{tabular} & \begin{tabular}[c]{@{}l@{}}Gain Information\\ (Information Capital)\end{tabular} \\ \hline
2. Cross-media Posting & Medium & 1 to Many & Single Post & Content Creation & \begin{tabular}[c]{@{}l@{}}Gain Information\\ (Information Capital)\end{tabular} \\ \hline
3. Clone Housekeeping & High & 1 to 1 & Non Verbal & Content Creation & \begin{tabular}[c]{@{}l@{}}Social Support\\ (Bonding Capital)\end{tabular} \\ \hline
4. Personalized Reachout & Low & 1 to 1 & Single Post & \begin{tabular}[c]{@{}l@{}}Make Friends\\ (Bridging Capital)\end{tabular} & \begin{tabular}[c]{@{}l@{}}Make Friends\\ (Bridging Capital)\end{tabular} \\ \hline
5. Undercover Rejections & Medium & 1 to 1 & Single Post & Content Creation & \begin{tabular}[c]{@{}l@{}}Social Support\\ (Bonding Capital)\end{tabular} \\ \hline
\begin{tabular}[c]{@{}l@{}}6. Post-completion for\\ Trending Topics\end{tabular} & Low & 1 to Many & Single Post & Content Creation & Entertainment \\ \hline
7. Mood Modifer & Medium & 1 to 1 & Conversation & \begin{tabular}[c]{@{}l@{}}Social Support\\ (Bonding Capital)\end{tabular} & \begin{tabular}[c]{@{}l@{}}Social Support\\ (Bonding Capital)\end{tabular} \\ \hline
8. Group-convo Simulation & Medium & Many to 1 & Conversation & \begin{tabular}[c]{@{}l@{}}Make Friends\\ (Bridging Capital)\end{tabular} & \begin{tabular}[c]{@{}l@{}}Gain Information\\ (Information Capital)\end{tabular}
\end{tabular}
\caption{Design Workbook Concepts Behaviors and Value}
\label{Tab:behavior}
\end{table}


\section{Interview Questions} \label{apx:questions}
\textbf{Phase 1 Target Interview Questions}
\begin{enumerate}
 \item Now that you’ve seen all the concepts, what are your general thoughts on the idea of AI clones in social media?
 \item Would you consider using AI clones for your own social media?
 \begin{enumerate}
     \item If so, how often would you see yourself using it?
 \end{enumerate}
 \item  In what ways can you see the clone enhancing your social experience?
 \item What are your major concerns of using such a technology?
 \item Was there a concept that really made a positive impression on you?
 \item What about one that you were concerned about or didn’t like?
 \item What is your ideal level of transparency?
 \item What is the minimum level of transparency for you to still use it?
 \item Who do you think should be responsible for disclosing AI clone use?
 \item What type of things do you think could go wrong?
 \item How would you react if something does go wrong?
 \item When something goes wrong, whose responsibility is it? Who should resolve it?
 \item Do you think the clone can influence your behavior?
 \item If you had your own clone, do you think you would behave more like how your clone behaves?
 \begin{enumerate}
     \item Why or why not?
 \end{enumerate}
 \item What kind of assumptions would you make about someone using clones?
 \item Would this depend on your prior impression of them?
 \item Do you think seeing the clone conversation of people you follow will affect your perception of them?
 \item Would you trust your clone to engage in these basic interactions?
 \item How would you react if someone’s clone said something offensive or rude to you?
 \item What impacts do you think having AI clones could have on you personally, and also on the general public as a whole?
\end{enumerate}

\noindent \textbf{Phase 2 Interactor Interview Questions}
\begin{enumerate}
 \item How did you feel interacting with a clone of [name]?
 \item What was your overall impression with the clones?
 \item What aspects of the clone are the most promising?
 \item What are some risks that the clone might have?
 \item Would you still use the clone despite these risks?
 \item How would you react and resolve the following breakdowns?
 \begin{enumerate}
      \item Mismatch between clone behavior and target
      \item Clone providing incorrect information
      \item Finding out that in interaction with the following person was with a clone:
      \begin{enumerate}
          \item Acquaintance
          \item Close friend
      \end{enumerate}
 \end{enumerate}
 \item From the following tradeoffs, which side do you value more and why?
 \begin{enumerate}
      \item Manual control vs. AI autonomy
      \item Realistic representation vs. Ideal representation
      \item Data privacy vs. Clone accuracy
      \item Transparency vs. Clone effectiveness
 \end{enumerate}
 \item Which concepts are more acceptable to you?
 \item Which concepts are you more open to disclosure?
 \item What type of relationship do you think the clones are best suited for?
 \item what type of relationships do you think it is acceptable to use clones in?
 \item What kind of assumptions would you make about someone using clones? 
 \item Do you think in general your impression of someone would change based on how their clone behaves?
 \item Do you value the AI interaction in of itself?
 \item What impacts do you think having AI clones could have on you personally, and also on the general public as a whole? (positive/negative)
\end{enumerate}

\noindent \textbf{Phase 3 Target Interview Questions}
\begin{enumerate}
 \item What was your overall impression with seeing your friends interact with your clone?
 \item How did you feel about the clone providing incorrect information to your friends?
 \item After seeing this conversation, do you feel like you would want to continue with talking to them with the same level of energy and friendliness?
 \item How would you feel if an acquaintance or stranger is able to talk to your clone and become closer to you without necessarily needing any reciprocation on your end?
 \item Where do you see the clones being the most useful for each of the social media platforms? 
 \item In terms of the concepts, which ones are more acceptable to you?
 \begin{enumerate}
     \item What makes them more acceptable, what makes the others less acceptable?
 \end{enumerate}
 \item Which concepts does disclosure matter the most for you and why?
 \item What type of relationship do you think the clones are best suited for clones?
 \item What role do you think AI clones should play in social media if any?
 \item What are some long term impacts that AI Clones may have online?
\end{enumerate}
\end{document}